\documentclass{icrc}

\usepackage{psfig,times}
\usepackage{graphicx} 

\begin{document}

\title{Stellar Sources of the Interstellar Medium}
\author[1] {F.-K. Thielemann} 
\author[1] {D. Argast} 
\author[1] {F. Brachwitz} 
\author[1] {G. Martinez-Pinedo}
\author[1] {R. Oechslin}
\author[1] {T. Rauscher}
\affil[1]{Dept. of Physics \& Astronomy, University of Basel,
Klingelbergstrasse 82, CH-4056 Basel, Switzerland}
\author[2] {W.R. Hix}
\author[2] {M. Liebend\"orfer} 
\author[2] {A. Mezzacappa}
\affil[2]{Physics Division, Oak Ridge National Laboratory,
Oak Ridge, TN 37831-6371, USA}
\author[3] {P. H\"oflich}
\affil[3]{Department of Astronomy, University of Texas,
Austin, TX 78712, USA}
\author[4] {K. Iwamoto}
\affil[4]{Department of Physics, Nihon University, Tokyo 101, Japan}
\author[5] {K. Nomoto}
\affil[5]{Department of Astronomy, University of Tokyo, Tokyo 113-033, Japan}
\author[6] {H. Schatz}
\affil[6]{National Superconducting Cyclotron Laboratory \& Department of Physics
and Astronomy, Michigan State University, East Lansing, MI 48824}
\author[7] {M.C. Wiescher}
\affil[7]{Department of Physics, University of Notre Dame, Notre Dame, 
IN 46556, USA}
\author[8] {K.-L. Kratz}
\author[8] {B. Pfeiffer}
\affil[8]{Institut f\"ur Kernchemie, Univ. Mainz, Fritz-Strassmann-Weg 2,
D-55099 Mainz, Germany}
\author[9] {S. Rosswog}
\affil[9]{Department of Physics and Astronomy, University of Leicester,
University Road, LE1 7RH, Leicester, UK}

\correspondence{fkt@quasar.physik.unibas.ch}

\firstpage{1}
\pubyear{2001}


\maketitle

\begin{abstract}
With the exception of the Big Bang, responsible for $^{1,2}$H, $^{3,4}$He,
and $^7$Li, stars act as sources for the composition of the interstellar
medium. Cosmic rays are related to the latter and very probably due to 
acceleration of the mixed interstellar medium by shock waves from supernova 
remnants. Thus, the understanding of the abundance evolution in 
the interstellar medium and especially the enrichment of heavy elements, as a 
function of space and time, is essential. It reflects the history of 
star formation and the lifetimes of the diverse contributing stellar objects. 
Therefore, the understanding of the endpoints of stellar evolution is 
essential as well. These are mainly planetary nebulae and type II/Ib/Ic 
supernovae as evolutionary endpoints of single stars, but also events in 
binary systems can contribute, like e.g.\ supernovae of type Ia, novae and 
possibly X-ray bursts and neutron star or neutron star - black hole mergers. 
Despite 
many efforts, a full and self-consistent understanding of supernovae (the main 
contributors to nucleosynthesis in galaxies) is not existing, yet.
Their fingerprints, however, seen either in spectra, lightcurves,
radioactivities/decay gam\-ma-rays or in galactic evolution, can help to
constrain the composition of their ejecta and related model uncertainties.
\end{abstract}

\section{Introduction}
The chemical composition of the interstellar medium in galaxies
acts as a witness of its stellar sources. The abundances found in the
solar system (shown in Fig.\ 1) are just a specific superposition of their 
yields, representing a snapshot in space and time.
Galactic evolution includes variations of the weigths in this superposition,
related to the evolutionary time scales of the individual contributing
stellar objects. It reflects also possible variations of the individual
yields as a function of metallicity.

Planetary nebulae, the endpoints of stars with $M$$<$8M$_\odot$, do contribute 
to light elements and heavy s-process nuclei
up to Pb and Bi. These are either products of H- and He-burning
(mainly $^4$He, $^{14}$N, $^{12}$C, [$^{16}$O, $^{22}$Ne]) or results of a
sequence of (slow) neutron captures and beta-decays (the s-process),
acting on pre-existing heavier nuclei in environments with small neutron
densities. The required neutrons are provided by 
$(\alpha,n)$-reactions in He-burning. 

\begin{figure}[t]
\psfig{file=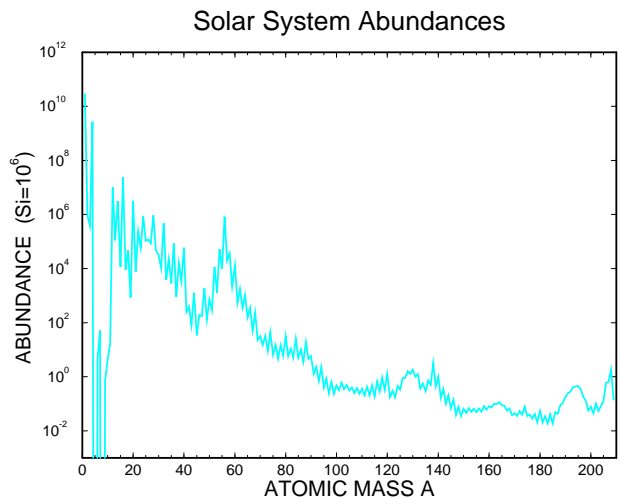,angle=0,width=9cm}
\caption{Abundances by number as found in the solar system
from meteorite samples and solar spectra (Grevesse \& Sauval 1998). Isotopic
abundances of different elements for the same mass number $A$ are added.
The units are scaled arbitrarily to a Si-abundance of Si of $10^6$.}
\end{figure}

The evolution of the elements in the range O through Ni
is dominated by two alternative explosive stellar sources, i.e.\ 
by the combined action of type II and type Ia supernovae (SNe II and Ia). 
SNe II, originating from massive stars ($M$$>$8M$_\odot$), leave a central 
neutron star and expell large amounts of the so-called alpha elements O, Ne, 
Mg, Si, S, Ar, Ca, Ti. Oxygen yields can be of the order 1M$_\odot$ or more  
while smaller amounts of Fe (and Fe-group elements like Cr, Mn, Fe, Co, Ni, 
Cu, Zn, Ga, Ge), typically about 0.1M$_\odot$, are produced in the innermost 
ejected mass zones. 
SNe Ia originate from those intermediate to low mass stars ($M$$<$8M$_\odot$)
which end as mass accreting white dwarfs in binary stellar systems 
with high mass accretion rates ($>$$10^{-8}$ M$_\odot$ y$^{-1}$).
In that case they experience stable hydrogen and helium shell burning of the
accreted matter, leading to a growing C/O white dwarf mass which approaches
the Chandrasekhar mass. This causes contraction and a complete disruption of 
the white dwarf after carbon is ignited in the center. Predominantly Fe-group 
elements are produced, on average of the order 0.6-0.8 M$_\odot$, and smaller 
amounts of Si, S, Ar, and Ca.

Novae, the other possible endpoint of mass accreting white dwarfs in binary
systems, occur for lower accretion rates. The unburned accreted hydrogen layer 
is ignited explosively. This causes only the ejection of the explosively 
processed hydrogen envelope of the order of $10^{-5}-10^{-4}$M$_\odot$ and
contributes relatively small amounts of matter to galactic evolution,
mainly $^{15}$N, $^{17}$O and a number of specific isotopes in the range Ne 
through S.
Type I X-ray bursts work in a similar way on accreting neutron stars in
binary systems. The envelope masses involved are even smaller ($\approx$
$10^{-9}$M$_\odot$)
and an ejection is not even clear due to the large gravitational field involved.
However, in case some matter can escape it would contain interesting products
from the rp-process (combined explosive hydrogen and helium burning with rapid
proton captures), including the
so-called light p-process isotopes of Se, Kr, Sr, Mo, and Ru.

The site for the production of the heaviest elements up to U and Th (the
r-process based on rapid neutron captures in environments with very high neutron
densities) is still debated and most probably related to type II supernovae 
and/or neutron star ejecta (from binary mergers or jets).

In the following sections 2 through 6 we want to present the status of 
understanding for most of
these objects, including especially supernova calculations, important
features in their progenitor evolution, the related nucleosynthesis and ejecta 
composition. Finally, in section 7, these stellar ejecta compositions  will be 
tested for their agreement with the chemical evolution of our Galaxy. 

\section{Stellar Evolution and Wind Ejecta}

H-burning converts $^1$H into $^4$He via pp-chains or the CNO-cycles. The 
simplest PPI chain is initiated by $^1$H(p,$e^+\nu$)$^2$H (p,$\gamma$)$^3$He and
completed by $^3$He($^3$He,2p)$^4$He. The dominant CNOI-cycle chain ,
$^{12}$C(p,$\gamma)^{13}$N($e^+\nu)$ $^{13}$C(p,$\gamma)^{14}$N(p,$\gamma)
^{15}$O ($e^+\nu)^{15}$N(p,$\alpha)^{12}$C, is controlled by 
$^{14}$N(p,$\gamma)^{15}$O, the slowest reaction of the cycle.
Further burning stages are
characterized by their major reactions, which are in 
He-burning $^4$He(2$\alpha,\gamma$)$^{12}$C (triple-alpha) and 
$^{12}$C($\alpha, \gamma$)$^{16}$O, in
C-burning $^{12}$C($^{12}$C, $\alpha$)$^{20}$Ne, and in
O-burning $^{16}$O($^{16}$O,$\alpha$)$^{28}$Si (e.g.\ Arnett \& Thielemann
1985, Thielemann \& Arnett 1985, Woosley \& Weaver 1995).
The alternative to fusion reactions are photodisintegrations which start to
play a role at sufficiently high temperatures when 30$kT$$\approx$$Q$ 
(the Q-value 
or energy release of the inverse capture reaction). This ensures the existence
of photons with energies $E_{\gamma}>$$Q$ in their Planck distribution and 
leads to Ne-Burning [$^{20}$Ne($\gamma,\alpha)^{16}$O, 
$^{20}$Ne($\alpha,\gamma)^{24}$Mg] at 
$T$$>$$1.5\times 10^9$K (preceding
O-burning) due to a small Q-value of $\approx$4~MeV and Si-burning 
at temperatures in excess of 3$\times$10$^9$K 
(initiated like Ne-burning by photodisintegrations, here of $^{28}$Si). 
The latter ends in a chemical equilibrium with an abundance 
distribution around Fe (nuclear statistical equilibrium, NSE)
(Hix \& Thielemann 1996, 1999). 

While this concept and the calculational details were unquestioned in wide
areas of the astrophysical community, the solar neutrino problem was still
shedding doubts on the quantitative understanding of stellar evolution.
The SNO experiment combined with SuperKamiokande data (Ahmad et al.\ 2001) gave
clear evidence for neutrino conversions, explaining the missing electron
neutrinos from the sun. This, together with constraints from helioseismology 
(Bahcall et al.\ 2001) gives strong support for our understanding of stellar
evolution. But it has to be realized that the role of convection and its
correct multi-D treatment (Asida \& Arnett 2000)
as well as stellar rotation, adding important features via diverse mixing 
processes (Heger et al.\ 2000ab, Meynet \& Maeder 2000), need further
refinement. Furthermore, 
the metallicity plays an essential role in stellar evolution and the amount of 
stellar mass loss (Langer et al.\ 1997, Charbonnel et al.\ 1999, Maeder 
\& Meynet 2000ab, 2001, Dominguez et al.\ 2001b).

\begin{figure}[t]
\psfig{file=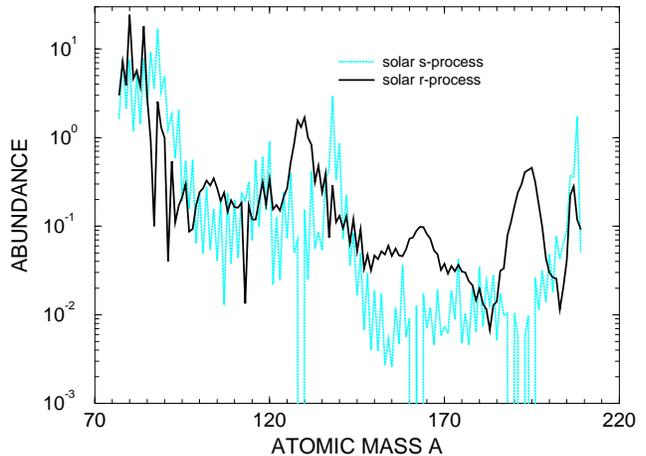,angle=0,width=9cm}
\caption{Decompositions of s-process and r-process abundances
(K\"appeler et al.\ 1989, K\"appeler 1999).
Notice the opposite behavior of the two processes with
respect to odd-even staggering as a function of mass number A.}
\end{figure}

Specific nuclear physics features enter during the latest stages of 
stellar evolution.
The high densities in late phases of O- and Si-burning
result in partially or fully degenerate electrons with increasing Fermi
energies (Nomoto \& Hashi\-moto 1988). When these supercede the Q-value 
thresholds of electron capture reactions, this allows for electron capture on 
an increasing number of initially Si-group (sd-shell) and later Fe-group 
(pf-shell) nuclei.
Because sd-shell reactions were well understood in the past (Fuller et al.\
1985), O-burning predictions were quite reliable. The recent progress in 
calculating pf-shell rates (Langanke \& Martinez-Pinedo 2000) led to drastic 
changes in the late phases of Si-burning (Heger et al.\ 2001ab).

Stars with masses $M$$>$8M$_\odot$ 
develop an onion-like composition structure, after passing through
all hydrostatic burning stages, and produce a collapsing core at the
end of their evolution, which proceeds to nuclear densities 
(Chieffi et al.\ 1998, Umeda et al.\ 2000, Heger et al.\ 2001ab).
The recent change in electron capture rates sets new
conditions for the Fe-core
collapse after Si-burning, the size of the Fe-core and its
electron fraction $Y_e$$=$$<$$Z/A$$>$ (Martinez-Pinedo et al.\ 2000)
which are important ingredients for a type II supernova explosion.

Less massive stars experience core and shell H- and He-burning and end as C/O
white dwarfs after strong episodes of mass loss (Hashimoto et al.\ 1993).
Their ejected nucleosynthesis yields have initially been predicted by Renzini
and Voli (1981). Recent reanalyses of their evolution and ejecta were undertaken
by  van den Hoek (1997) and Charbonnel et al.\ (1999).
Shell He-burning is also identified as the source of the s-process,
where $(\alpha,$n)-reactions, induced by the He-fuel, provide neutrons
for sequences of neutron captures and beta-decays of unstable nuclei.
$^{22}$Ne($\alpha$,n) is a natural neutron source, with $^{22}$Ne - originating
from $^{14}$N left over from H-burning - via 
$^{14}$N($\alpha,\gamma)^{18}$F$(e^+ \nu_e)^{18}$O($\alpha,\gamma)^{22}$Ne.
Details of mixing processes play a role in providing the stronger
neutron source $^{13}$C($\alpha$,n), which requires the mixing of protons
(hydrogen) into He-burning layers, leading to 
$^{12}$C(p,$\gamma)^{13}$N ($e^+\nu_e)^{13}$C (K\"appeler 1999,
Busso et al.\ 2001, Van Eck et al.\ 2001).

More details about the production of heavy elements via slow and rapid
neutron capture are given in section 6. Fig.\ 2 shows the decomposition
of the solar abundances of heavy nuclei into the s- and r-process components
(from K\"appeler et al.\ 1989, K\"appeler 1999).

Above, we discussed the "conventional" endstages of single star evolution with
moderate progenitor masses, which dominate galactic nucleosynthesis. However,
we know that above a (not yet well determined) limit in the range 25-60M$_\odot$
massive stars end with an Fe-core which is too large for a successful
supernova explosion. This causes the central collapse to proceed to a 
stellar mass black hole. Dependent on a pending understanding of
magneto-hydrodynamic effects, accretion of the remaining stellar envelope
onto these black holes can lead to jets and "hypernova" explosions
(MacFadyen \& Woosley 1999, Cameron 2001ab, MacFadyen et al.\ 
2001). The latter are possibly related to one class of gamma-ray bursts.
Their nucleosynthesis has been discussed by Nakamura et al.\ (2001).

Very massive stars with several 100 M$_\odot$, which experience contraction
due to the pair instability at high temperatures, undergo a complete disruption
by explosive nuclear oxygen burning (VMOs, Heger \& Woosley 2002). They should 
be considered in a similar way as hypernovae. The understanding of the initial 
mass function, describing the formation frequency as a function of progenitor
mass, is, however, still marginal.

\section{Type II Supernovae}

The core collapse in massive stars, initiated via pressure reduction due
to electron captures at the end of their stable evolution of hydrostatic
core burning stages, leads to a bounce at nuclear densities and the formation
of a hot neutron star which gained gravitational binding energy of the order 
$10^{53}$erg (Bethe 1990). This provides neutrinos, diffusing out fast in 
comparison to other particles due to their small interaction cross sections.
This was observed for SN 1987A (Burrows 1990).
Neutrinos can deposit their energy via 
$\nu_e +n\rightarrow p+e^-$ and $\bar\nu_e +p\rightarrow n+e^+$ in adjacent
layers. An energy deposition exceeding $10^{51}$erg would be sufficient to
cause an explosion and ejection of all layers surrounding the proto neutron
star. The present situation in supernova modeling is that self-con\-sistent 
spherically-symmetric calculations 
(with the presently known microphysics) do not yield successful explosions
based on neutrino energy deposition from the hot collapsed central core 
(neutron star) into the adjacent layers (Rampp \& Janka 2000). Even 
improvements in neutrino transport, solving the full Boltzmann transport
equation for all neutrino flavors (Mezzacappa et al.\ 2001), and a fully
general relativistic treatment (Liebend\"orfer et al.\ 2001ab, Liebend\"orfer 
et al.\ 2002) did not change this situation (see Figs.3 and 5). 
This seems to be the same for multi-D calculations (e.g.\ Mezzacappa et al.\
1998), which however lack good neutrino transport schemes
and do not yet consider the possible combined action of rotation and magnetic 
fields.

\begin{figure*}[t]
\figbox{}{}{\psfig{file=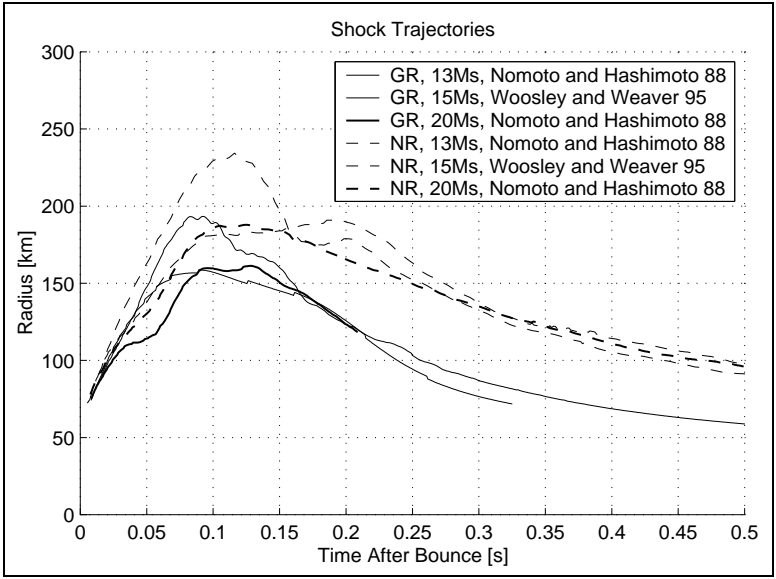,width=10cm}}
\caption{Results of core collapse supernova simulations from Liebend\"orfer
et al.\ (2002) for a variety of
massive stellar models from Nomoto \& Hashimoto (1988) and Woosley \& Weaver
(1995). Shown is the position of the shock as a function of time after central
core bounce at nulear densities. We see that the shocks are strongest for
the least massive stars. But all of them stall, recede and turn into accretion
shocks, i.e. not causing successfull supernova explosions. For these 
(non-successful) simulations, the full general
relativistic treatment (GR) weakens the shock further in comparison to
the non-relativistic treatment (NR).}
\end{figure*}

\begin{figure}
\vspace{-0.5cm}
\centerline{\psfig{file=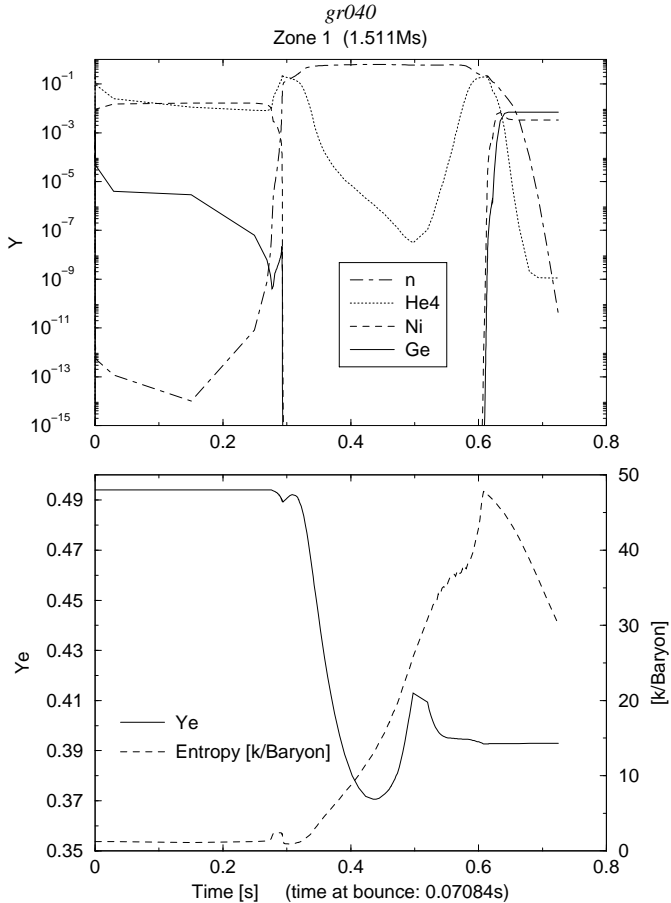,width=9cm}}
\caption{
Hydrodynamic simulations with varied (reduced) neutrino opacities
(scattering cross sections on nucleons reduced by 60\%)
lead to larger neutrino luminosities and make successful supernova explosions
possible (Hauser et al.\ 2002). Here we see the $Y_e$ and entropy evolution
of the innermost ejected zone of a 20M$_\odot$ star after core collapse,
central bounce and neutrino heating (buttom). The top figure provides the
composition evolution.}
\label{pascal}
\end{figure}

\begin{figure*}
\vspace{-1cm}
\centerline{\psfig{file=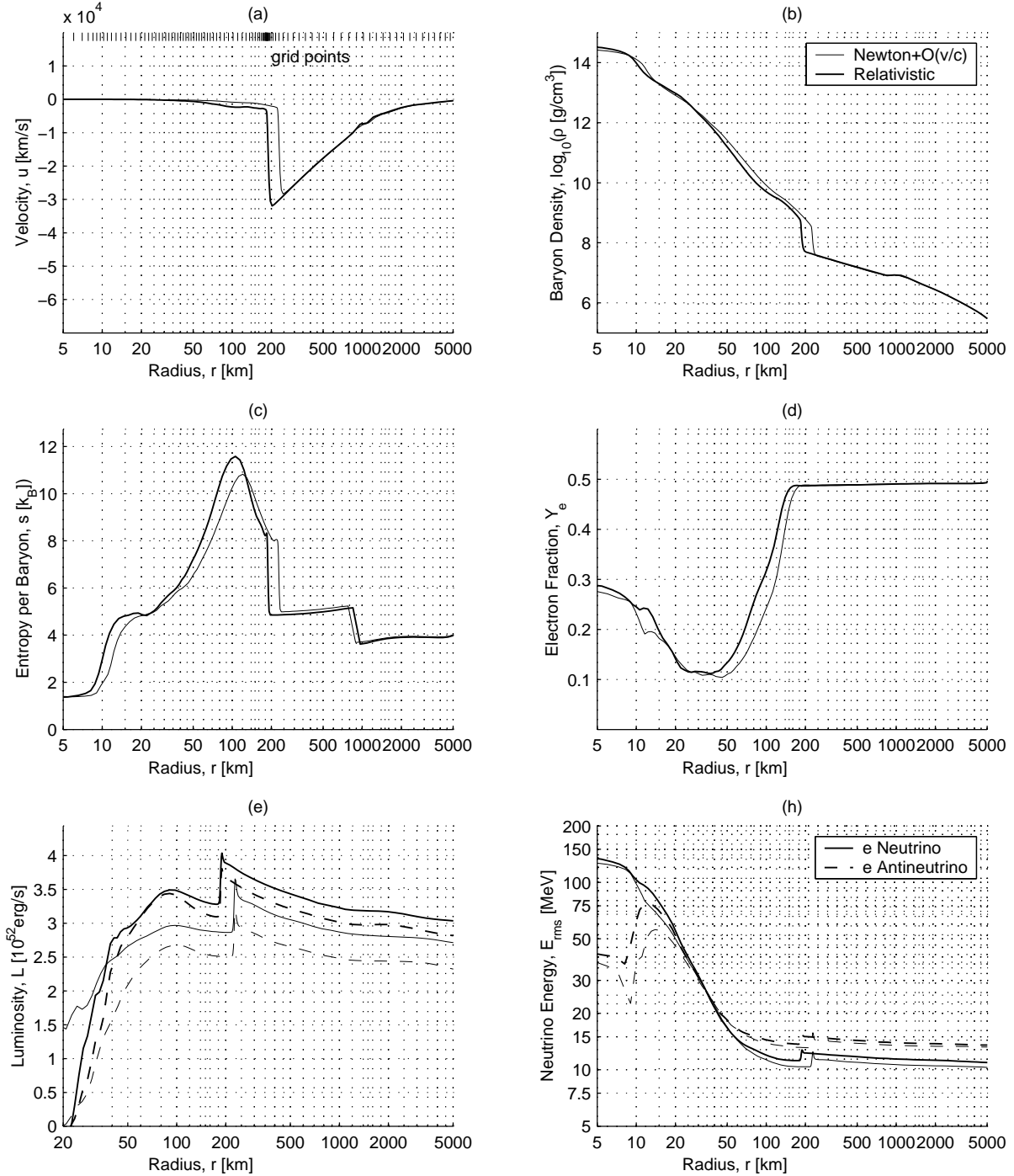,width=16cm,angle=0}}
\caption{Details from a core collapse and explosion simulation 
(Liebend\"orfer et al.\ 2001a)
of a 13$M_\odot$ stellar model from Nomoto \& Hashimoto (1988) 100ms after the 
central bounce at nuclear densities in the phase of neutrino heating via 
$\nu_e +n\rightarrow p+e^-$ and $\bar\nu_e +p\rightarrow n+e^+$. A 
Newtonian hydro calculation with Boltzmann neutrino transport and 
terms up to the order $v/c$ is compared to a fully general relativistic 
treatment in hydro and neutrino transport. Shown are (a) the velocity profile 
and the location of 
the adaptive grid points, (b) the rest mass density profile, (c) the entropy,
(d) the electron fraction $Y_e$, the net electron (or proton) to nucleon
ratio, i.e. a measure of the neutron-richness of matter, (e) the neutrino
luminosity, and (f) the average energy of electron neutrinos and 
anti-neutrinos. We see that the shock is located at about 200km and at this
point in time (100ms after bounce) has the features of an accretion shock 
(no positive 
velocities) rather than an outward moving shock. This underlines the negative
outcome of spherically symmetric models with presently known microphysics.}
\end{figure*}

\begin{figure*}
\centerline{\psfig{file=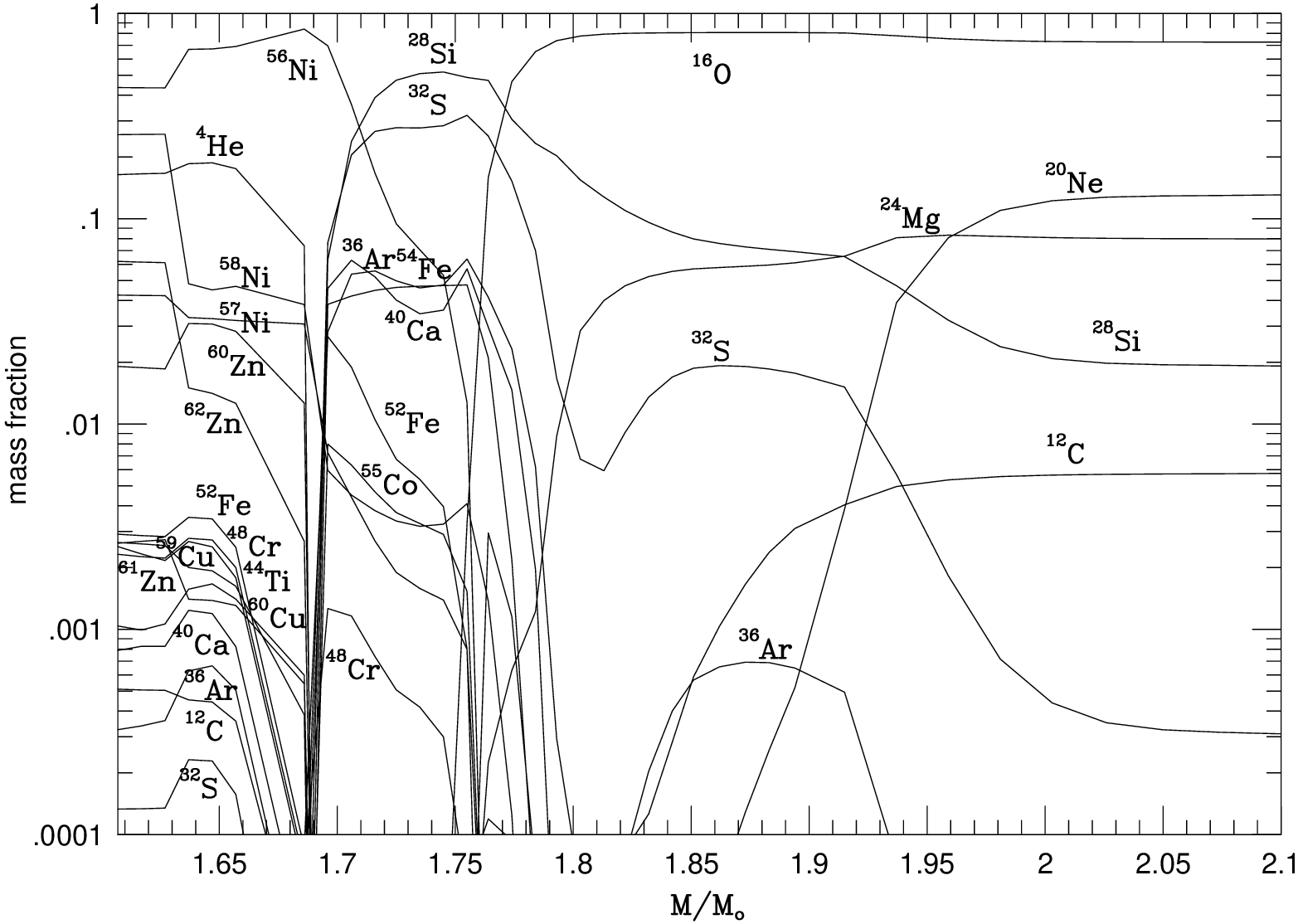,width=10cm}}
\vspace*{-4.2cm}
\caption{Isotopic composition for the explosive C-, Ne-, O- and Si-burning
layers of a core collapse
supernova from a 20M$_\odot$ progenitor star with a 6M$_\odot$
He-core and an induced net explosion energy of $10^{51}$ erg, remaining in 
kinetic energy of the ejecta (from Thielemann et al.\ 1996). $M(r)$ indicates 
the radially enclosed mass, integrated from the stellar center.
The exact mass cut in $M(r)$ between neutron star and
ejecta and the entropy and $Y_e$ in the innermost ejected layers
depend on the details of the (still open) explosion mechanism.
The abundances of O, Ne, Mg, Si, S, Ar, and Ca dominate strongly over
Fe (decay product of $^{56}$Ni), if the mass cut is adjusted to 0.07M$_\odot$
of $^{56}$Ni ejecta as observed in SN 1987A.
}
\end{figure*}

The hope that the neutrino driven explosion mechanism could still succeed
is based on uncertainties which affect neutrino luminosities (neutrino opacities
with nucleons and nuclei, Prakash et al.\ 2001), convection in the hot 
proto-neutron star (Keil \& Janka 1995),
as well as the efficiency of neutrino energy deposition (convection in the
adjacent layers, Wilson \& Mayle 1993).
Fig.\ 4 shows the temporal $Y_e$ and entropy evolution of the innermost zone of 
a 20 M$_\odot$ SN II simulation (Hauser et al.\ 2002), based on 
Lieben\-d\"orfer et al.\ 
(2001a), but with varied neutrino opacities (reducing electron neutrino and 
antineutron scattering cross sections on nucleons by 60\%) which permits a 
successful delayed explosion. This is only a parameter study, but it shows
at least how options which increase the neutrino luminosity may lead to
success.

Observations show typical kinetic energies of $10^{51}$~erg in supernova
remnants. Thus, even without a correct understanding of the explosion
mechanism, this permits one to perform light curve as well as explosive 
nucleosynthesis calculations by introducing a shock of 
appropriate energy in the pre-collapse stellar model 
(Woosley \& Weaver 1986, Thielemann et al.\ 1990, Aufderheide et al.\ 1991,
Woosley \& Weaver 1995, Thielemann et al.\ 1996, Nomoto et al.\ 1997, 
Hoffman et al.\ 1999,
Nakamura et al.\ 1999, Umeda et al.\ 2000, Rauscher et al.\ 2001a, 2002).
Such induced calculations lack 
self-consistency and cannot predict the
ejected $^{56}$Ni-masses from the innermost
explosive Si-burning layers (powering supernova light curves
by the decay chain $^{56}$Ni-$^{56}$Co-$^{56}$Fe) due to missing
knowledge about the detailed explosion mechanism and therefore the mass cut 
between the neutron star and supernova ejecta.
However, the intermediate mass elements Si-Ca are only dependent
on the explosion energy and the stellar structure
of the progenitor star, while abundances for elements like O and Mg are 
essentially determined by the stellar progenitor evolution.
Thus, when moving in from the outermost to the innermost ejecta of a SN II
explosion, we see an increase in the complexity of our understanding,
depending (a) only on
stellar evolution, (b) on stellar evolution and explosion energy, and
(c) on stellar evolution and the complete explosion mechanism (see Fig.\ 6).

The possible complexity of the explosion mechanism, including multi-D effects,
does not affect this (spherically symmetric) discussion of explosive 
nucleosynthesis severely. The 2D-calculations of Kifionidis et al.\ (2000)
show a spherically symmetric shock front after the explosion is initiated,
leading to spherical symmetry in explosive nuclear burning when passing
through the stellar layers. Only after the passage of the shock front,
the related temperature decline and freeze-out of nuclear reactions, the
final nucleosynthesis products can be distributed in non-spherical geometries
due to mixing by hydrodynamic instabilities. Thus, the total mass of
nucleosynthesis yields shown in Fig.\ 6 is not changed, only its geometric
distribution.

The correct prediction of
the amount of Fe-group nuclei ejected (which includes also one of the 
so-called alpha elements, i.e.\ Ti) and their relative composition
depends directly on the explosion mechanism and the size 
of the collapsing Fe-core. 
Three types of uncertainties are inherent in the Fe-group ejecta,
related to (i) the total amount of Fe(group) nuclei ejected and the
mass cut between neutron star and ejecta, mostly
measured by $^{56}$Ni decaying to $^{56}$Fe, (ii) the total explosion energy
which influences the entropy of the ejecta and with it the amount of
radioactive $^{44}$Ti as well as $^{48}$Cr, the latter decaying later to
$^{48}$Ti and being responsible for elemental Ti, and (iii) finally
the neutron richness or $Y_e$=$<Z/A>$ of the ejecta,
dependent on stellar structure, electron captures and neutrino interactions.
$Y_e$ influences strongly the ratios of isotopes 57/56 in
Ni(Co,Fe) and the overall elemental Ni/Fe ratio, the latter being dominated
by $^{58}$Ni and $^{56}$Fe. 

The pending understanding of the explosion mechanism also affects possible
r-process yields for SNe II
(Takahashi et al.\ 1994, Woosley et al.\ 1994, Qian \& Woosley 1996,
Freiburghaus et al.\ 1999a, McLaughlin et al.\ 1999,
Nagataki \& Kohri 2001, Thompson et al.\ 2001, Wanajo et al.\ 2001,
Sumiyoshi et al.\ 2001). A more detailed discussion
is given in section 6.

\section{Type Ia Supernovae}

\begin{figure*}[t]
\centerline{\psfig{file=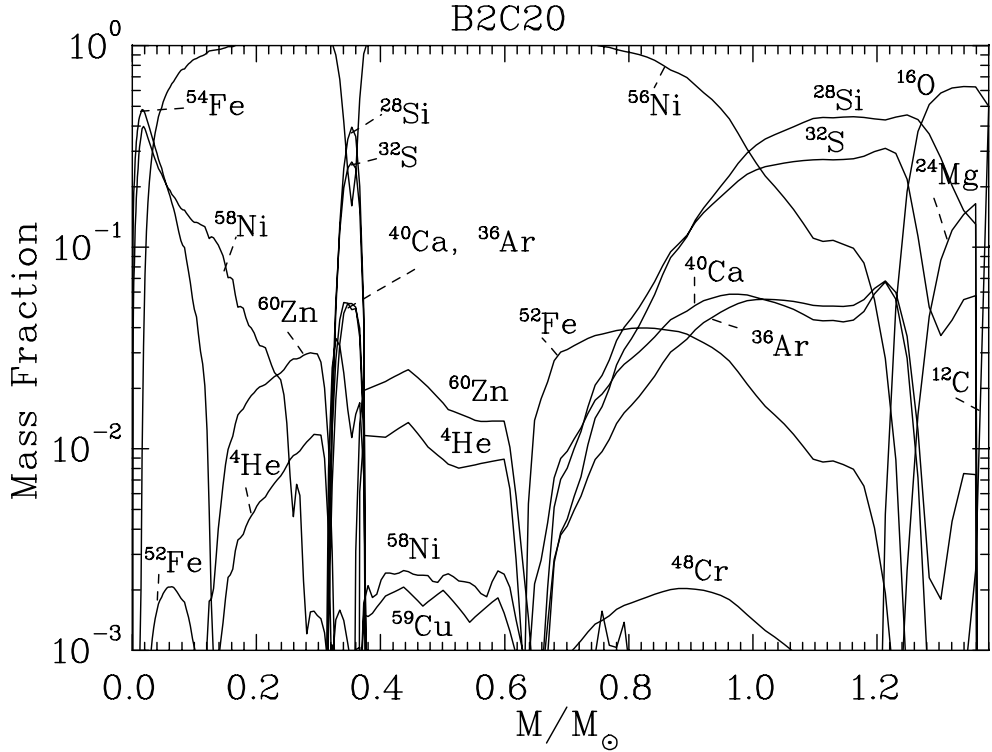,angle=90,width=13cm}}
\caption{Isotopic composition for the layers of a type Ia supernova,
starting thermonuclear burning with a deflagration front which turns
into a detonation at 0.32 M$_\odot$. This is shown in the $^{56}$Ni 
feature which sandwiches explosive O-burning products like $^{28}$Si through 
$^{40}$Ca.
$M(r)$ indicates the radially enclosed mass, integrated
from the stellar center. We see the products of explosive Si-burning 
($^{56}$Ni), O-burning ($^{28}$Si), Ne-burning ($^{16}$O and $^{24}$Mg),
minor amounts of C-burning ($^{20}$Ne) and unburned matter at the surface.
The central Fe-group composition depends on $Y_e$ which is directly related
to the amount of electron capture on free protons and nuclei.
}
\end{figure*}

 There are strong observational and theoretical indications that
SNe Ia are thermonuclear explosions of accreting white dwarfs in binary
stellar systems (H\"oflich \& Khokhlov 1996, Nugent et al.\ 1997, Nomoto et al.\
2000, Livio 2001) with central carbon ignition leading to a thermonuclear 
runaway, because the pressure is dominated by a degenerate electron gas and
shows no temperature dependence. This prevents a stable and controlled burning,
causing a complete explosive
disruption of the white dwarf (Nomoto et al.\ 1984, Woosley \& Weaver 1994). 
The mass accretion rates determine the ignition densities.  A flame front
then propagates at a subsonic speed as a deflagration wave due
to heat transport across the front (Hillebrandt \& Niemeyer 2000).

The averaged spherical flame speed 
depends on the development of instabilities on various scales at the
flame front. (The flame front thickness is of the order 10$^{-4}$cm and
determined by heat conduction via the electron mean free path.)
Multi-dimensional hydro simulations suggest a speed
$v_{\rm def}$ as slow as a few percent of the sound speed $v_{\rm s}$
in the central region of the white dwarf.  
Electron capture affects the central electron fraction $Y_e$
which depends on 
(i) the electron capture rates of nuclei,
(ii) $v_{\rm def}$,
influencing the time duration of matter at high temperatures (and with
it the availability of free protons for electron captures), and (iii) the 
central
density of the white dwarf $\rho_{ign}$
(increasing the electron chemical potential i.e.\ the Fermi energy)  
(Iwamoto et al.\ 1999, Brachwitz et al.\ 2000, Langanke \& Martinez-Pinedo 2000).
After an initial deflagration in the central layers, the
deflagration might turn into a detonation (supersonic burning front)
at larger radii and lower densities (Niemeyer 1999). The nucleosynthesis 
consequences can be viewed in Fig.\ 7 (Brachwitz et al.\ 2002).

Nucleosynthesis constraints can help to find the "average" SN Ia
conditions responsible for their contribution to
galactic evolution, i.e.\ especially the Fe-group composition
(Thielemann et al.\ 1986).
Ignition densities $\rho_{ign}$ dominate the very central amount of
electron capture and thus $Y_e$. The deflagration speed $v_{\rm def}$
affects the time duration of burning in a zone and with it the possible
amount of electron captures on free protons and nuclei. It is also responsible
for the time delay between the arrival of the information that a burning front
is approaching (the information is propagating with sound speed and causes
expansion of the outer layers) and the actual arrival of the burning front.
Burning at lower densities causes less electron captures. Thus, $v_{\rm def}$
determines the resulting $Y_e$ -gradient as a function of radius 
(Iwamoto et al.\ 1999).
$Y_e$ values of 0.47-0.485 lead to dominant abundances of
$^{54}$Fe and $^{58}$Ni, values between 0.46 and 0.47 produce
dominantly $^{56}$Fe, values in the range of 0.45 and below are
responsible for $^{58}$Fe, $^{54}$Cr, $^{50}$Ti, $^{64}$Ni, and values
below 0.43-0.42 are responsible for $^{48}$Ca.  
The intermediate $Y_e$-values 0.47-0.485 exist in all cases, but the masses
encountered which experience these conditions depend on the $Y_e$-gradient
and thus $v_{def}$. Whether the lower values with $Y_e$$<$0.45 are attained,
depends on the central ignition density $\rho_{ign}$. Therefore, 
$^{54}$Fe and $^{58}$Ni are indicators of $v_{def}$ while  $^{58}$Fe, 
$^{54}$Cr, $^{50}$Ti, $^{64}$Ni, and $^{48}$Ca are a measure of $\rho_{ign}$.
A test for these  (hydrodynamic) model parameters is shown in Fig.\ 8 where
B1 and B2 indicate increasing propagation speeds of the burning front and C20 
through C80 increasing ignition densities (results from Brachwitz et al.\
2002). 

\begin{figure}
\centerline{\psfig{file=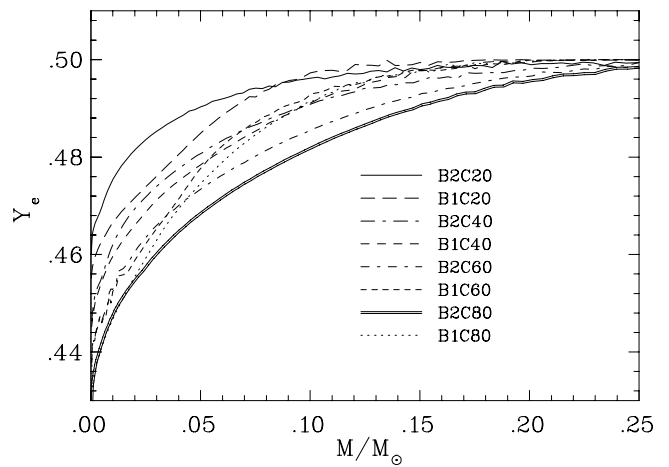,angle=90,width=8.5cm}}
\caption{$Y_e$ after freeze-out of nuclear reactions
measures the electron captures on free protons and
nuclei. Small
burning front velocities lead to steep $Y_e$-gradients which flatten
with increasing velocities (see the series of B1 vs.~the B2 models).
Lower central ignition densities shift the curves up (see changes from 
C20 through C80, i.e. central ignitions at 2-8$\times 10^9$ g cm$^{-3}$), but 
the gradient is the same for the same propagation speed (from Brachwitz et 
al.~2002). 
}
 \end{figure}

\begin{figure*}[t]
\centerline{\psfig{file=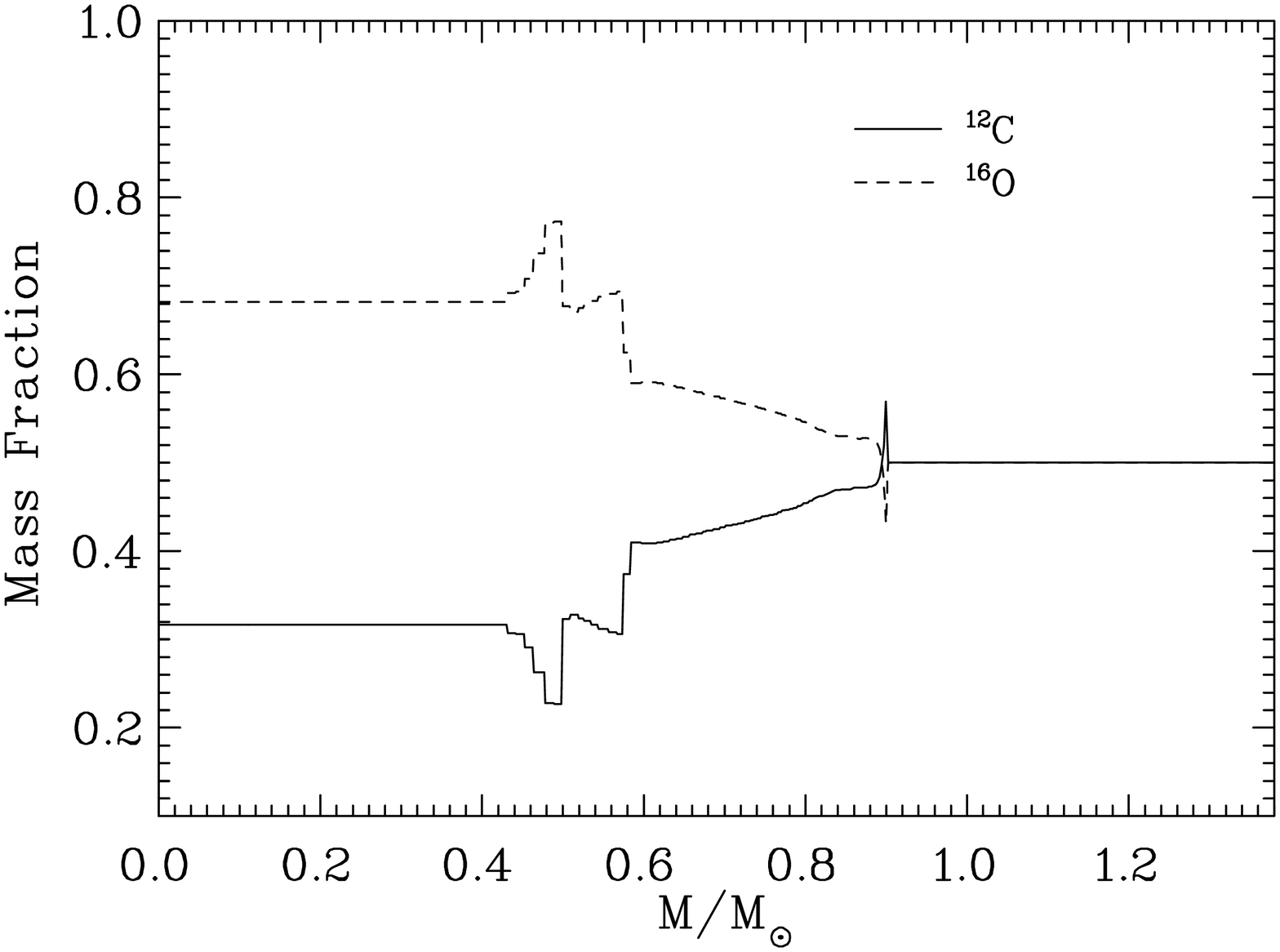,angle=90,width=8.5cm}
\hspace{0.5cm}
\psfig{file=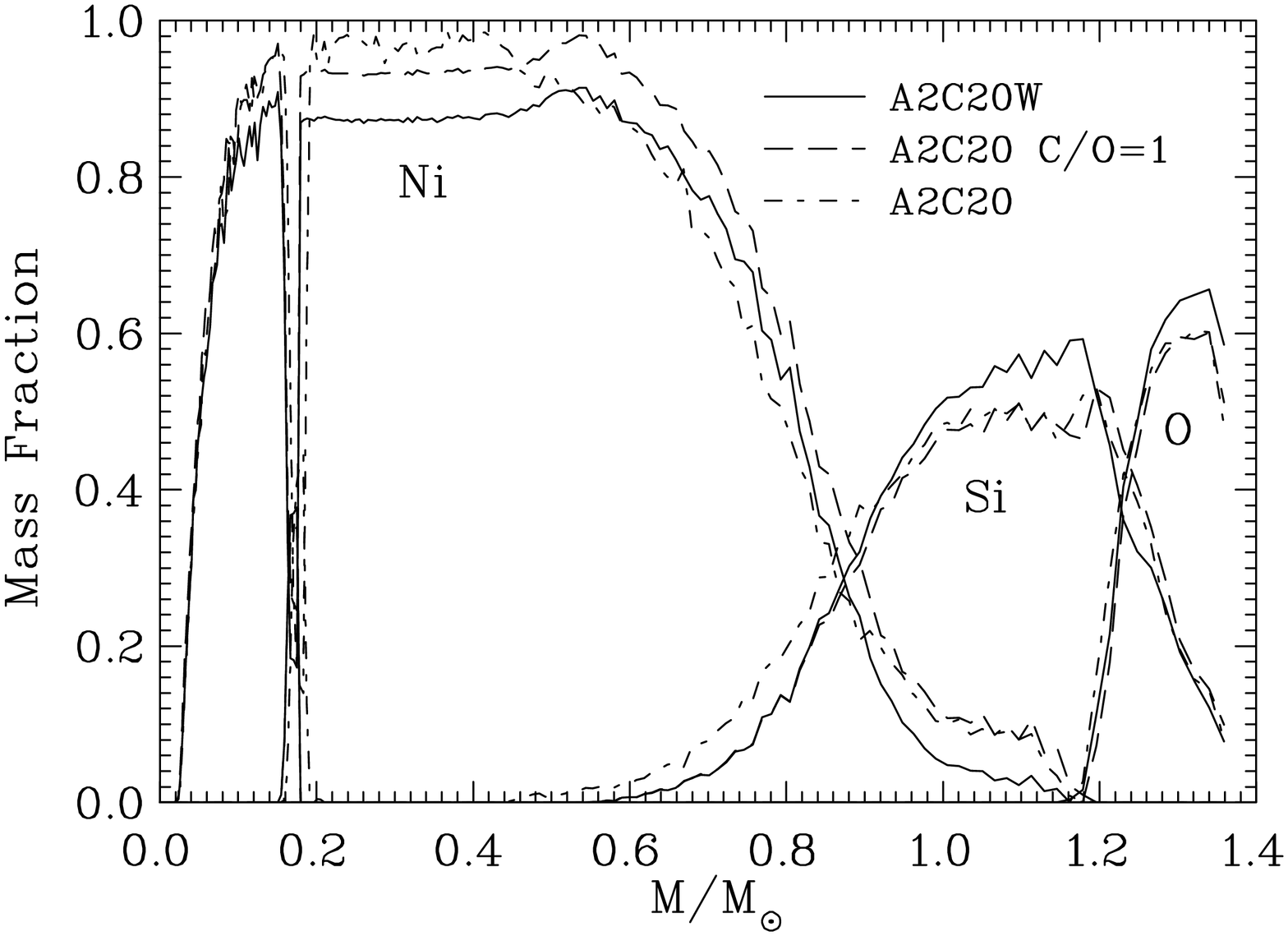,angle=90,width=8.5cm}}
\caption{Left: Mass fractions of C and O throughout the white dwarf before 
undergoing
thermonuclear ignition. The inner part (where O dominates) shows the
initial white dwarf before accretion, the outer part the matter which
underwent H- and He-burning during the accretion phase (at higher temperatures)
where the mass fractions of C and O are comparable.
Right: Mass fractions of Ni ($^{56}$Ni), Si ($^{28}$Si) and O 
($^{16}$O) throughout the white dwarf after a thermonuclear explosion.
The model A2C20 stands for a burning front propagation of 2\% of the
sound speed and a central ignition density of $2\times 10^9$ g cm$^{-3}$ and
is based on a white dwarf composition as indicated on the left.
C/O=1 stands for a modified white dwarf composition with equal amounts of
C and O also in the inner region of the original white dwarf. The third
model A2C20W reflects a higher metallicity of the material. This is 
measured by the amount of $^{22}$Ne (2.5\% in comparison to 1\% for
the other models, Brachwitz et al.\ 2002).
}
 \end{figure*}

\begin{figure*}
\centerline
{\psfig{file=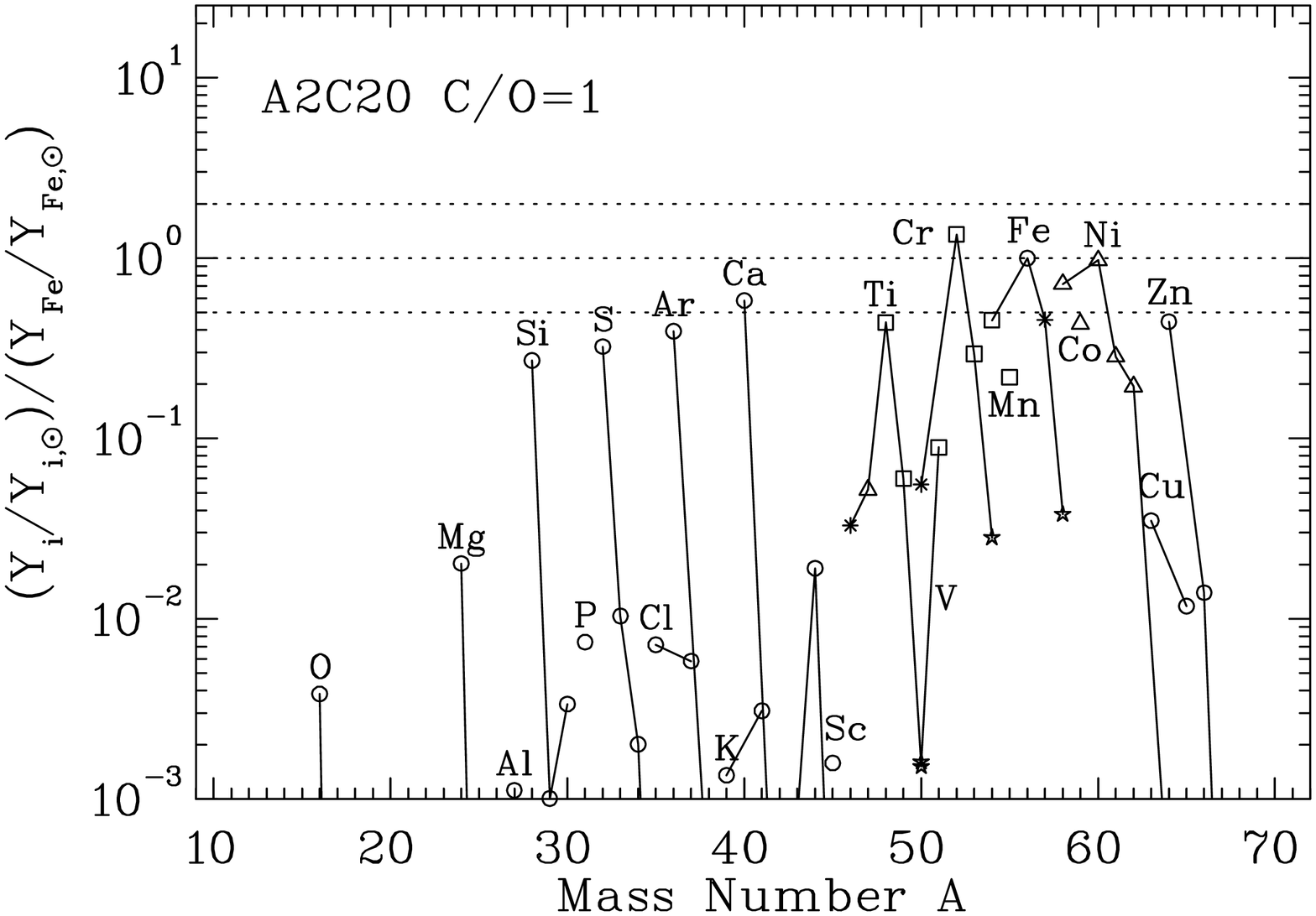,angle=90,width=8.5cm}
\hspace{0.5cm}
\psfig{file=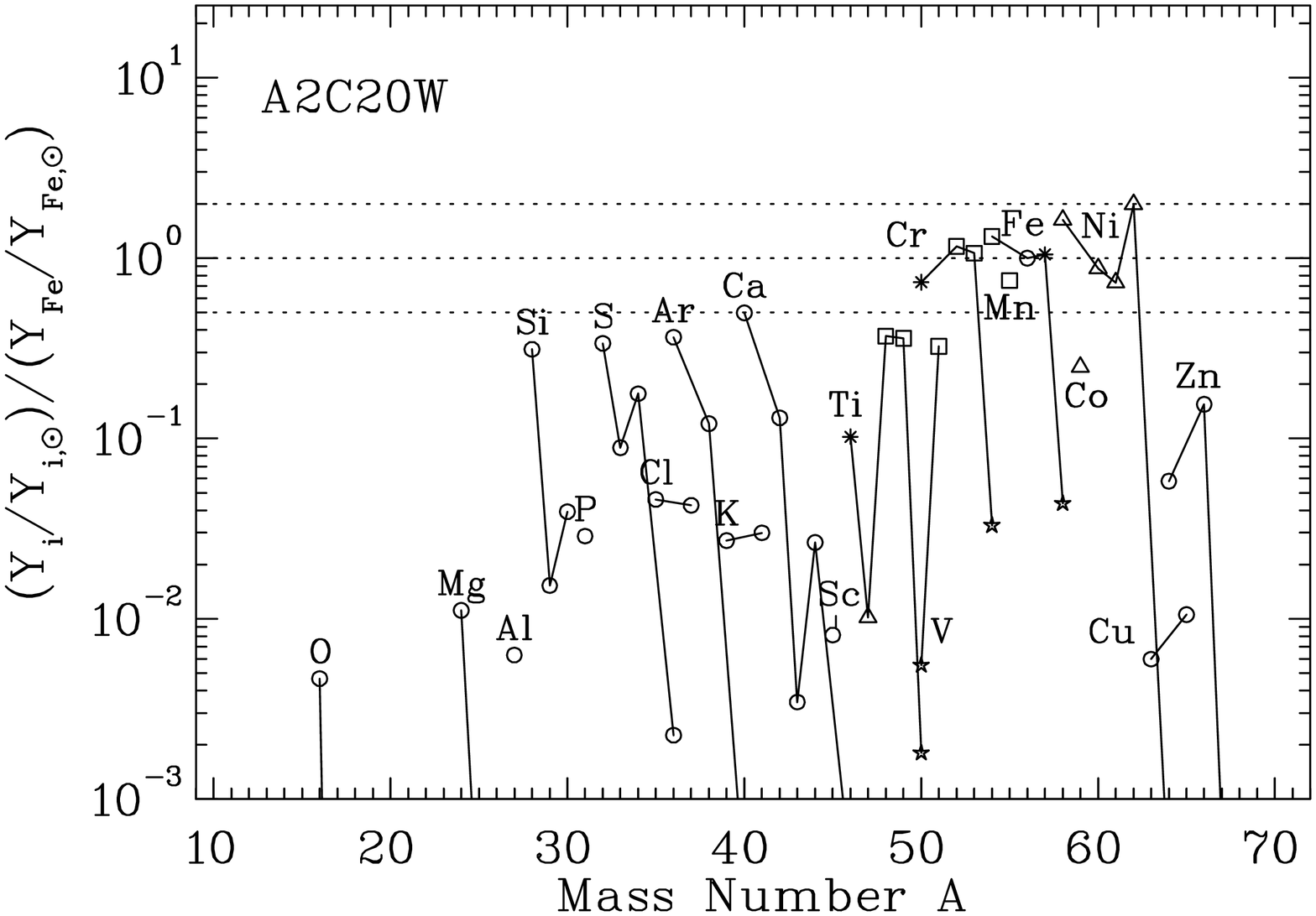,angle=90,width=8.5cm}}
\caption{Abundance ratios of nuclei compared to their solar values
and normalized for $^{56}$Fe, the decay product of  $^{56}$Ni. 
Isotopes of the same element are connected by lines. The general
feature is that the ratio of Fe to Si-Ca is about a factor of 2-3 higher than 
solar. For the changes in the Fe-group composition see text (from
Brachwitz et al.\ 2002).
}
 \end{figure*}

Nuclear uncertainties based on electron capture rates enter as well
(Brachwitz et al.\ 2000, 2002). 
Conclusions from these results are:
(i) a $v_{def}$ in the range 1.5--3\% of the sound speed is preferred
(Iwamoto et al.\ 1999), and (ii) the change in 
electron capture rates (Langanke \& Martinez-Pinedo 2000) made it possible to 
have ignition densities as high as $\rho_{ign}$$=$$2\times 10^{9}$ g cm$^{-3}$
(as expexted from typical accretion rates, see e.g.\ Dominguez et al.\ 2001a)
without destroying the agreement
with solar abundances of very neutron-rich species (Brachwitz et al.\ 2000).
It seems, however, hard to produce amounts of $^{48}$Ca sufficient to
explain solar abundances from SNe Ia (in spherical symmetry) when applying 
more realistic electron 
capture rates, even for very high ignition densities
(Woos\-ley 1997, Brachwitz et al.\ 2002).

While uncertainties in the detailed 3D burning front propagation have
been parametrized in terms of propagation speed and detonation transition
density, it is expected that in a self-consistant simulation both "parameters" 
will adjust themselves  to unique values for the
same initial models. The ignition density is a system indicator, reflecting
the accretion history in a binary system. A similar hidden system parameter
is the mass of the initial white dwarf, which determines the C/O ratio
within the original white dwarf before accretion sets in (H\"oflich 2000,
Dominguez et al.\ 2001a and Fig.\ 9 left). The accretion
burning at higher temperatures leads always to comparable C and O mass 
fractions. 

Fig.\ 9 (right) shows the resulting abundances for explosions with different
C/O ratios also in the central part of the initial white dwarf (C and O 
like in Fig.\ 9 left or C/O=1).
The reason is related to the different energy release for burning $^{16}$O
or $^{12}$C to $^{56}$Ni. A further change is seen if the metallicity of
the object (or the accreted matter) is changed. Existing CNO burns in 
H-burning to $^{14}$N and in He-burning to $^{22}$Ne, a nucleus with two more
neutrons than protons. This affects the energy generation in the outer
Si-burning layers, where the neutron-richness of matter is not due to
electron captures as in the inner high-density layers but originating
from the initial pre-ignition composition (measured here in the fraction
of $^{22}$Ne, 1 vs. 2.5\% in case W). We see how the production of
$^{56}$Ni is changed which directly determines the light curve
properties. The central C/O ratio, reflecting the white dwarf mass
in the pre-explosion binary system (Domiguez et al.\ 2001a), can thus serve as 
a parameter which determines the supernova and lightcurve properties and
must be related to the empiricial brightness/decline relation
(Riess et al.\ 2000, Leibundgut 2001). In Fig.\ 9 we see, however, also that
the initial metallicity (measured by the original $^{22}$Ne content)
can have similar effects. This introduces an empirical two rather than one 
parameter dependence (see also H\"oflich et al.\ 1998). However, detailed
analyses by Dominguez et al.\ (2001a) have shown that the latter effect is
smaller than the observational changes towards high redshifts. This supports
the cosmological use of SNe Ia also at high redshifts.

The general nucleosynthesis outcome of SNe Ia is dominated by Fe-group
products, but involves sizable fractions of Si-Ca and minor amounts of
unburned or pure C-burning products (e.g.\ C, O, Ne, Mg). The ratio of
Fe to Si-Ca is about a factor of 2-3 higher than in solar composition
(Greves\-se \& Sauval 1998). This typical feature is shown in Fig.\ 10 for
models discussed above. In principle one expects major differences for the
Fe-group composition as a function of parameters like burning front
propagation and ignition density (Iwamoto et al.\ 1999, Brachwitz et al.\
2000). Here we present models which agree with the general constraints
but also show a change in metallicity. One difference which becomes
apparent immediately is the change in Mn. $^{55}$Mn (the only stable isotope
of Mn) is a decay product of $^{55}$Co which is mainly produced in 
incomplete Si-burning. In this respect it was discussed by Iwamoto et al.\
(1999) as an indicator of the deflagration-detonation transition in
delayed detonation models. Here we see that it is affected by metallicity
as well (strongly changing from A2C20 to A2C20W, more details in Brachwitz
et al.\ 2002).

\section{Novae and X-ray Bursts}

\subsection{Novae and explosive hydrogen burning}

The major astrophysical events which involve explosive H-burning are novae
and type I X-ray bursts. Both are related to binary stellar systems with
hydrogen accretion from a binary companion onto a compact object at small
accretion rates which leave the H/He-fuel unburned. In the case of novae the 
compact object is a white dwarf, in the case of X-ray bursts it is a neutron
star. Ignition sets in
when a critical layer mass $\Delta M$ is surpassed, in novae (dependent on
the white dwarf mass) of the order $10^{-5}-10^{-4}$M$_\odot$, in type I
X-ray bursts of the order $10^{-9}$M$_\odot$. This leads to
explosive ignition of the accreted H-layer under degenerate
conditions (similar to central C-ignition in SNe Ia). For high densities
the pressure is dominated by the degenerate electron gas and shows no 
temperature dependence. This
prevents a stable and controlled burning, leading therefore to a 
thermonuclear runaway. While the ignition is always based on pp-reactions
(as in solar hydrogen burning), the runaway leads to the hot CNO-cycle
$^{12}$C(p,$\gamma)^{13}$N(p,$\gamma)^{14}$O($e^+\nu)^{14}$N(p,$\gamma)^
{15}$O($e^+\nu)$ 

\noindent $^{15}$N(p,$\alpha)$
$^{12}$C, branching out partially via
$^{15}$N(p,$\gamma)^{16}$O(p,$\gamma)$ $^{17}$F($e^+\nu)$
$^{17}$O(p,$\gamma)^{18}$F...

\begin{figure}
\vspace{-3cm}
\centerline{\hspace{1cm}
\psfig{file=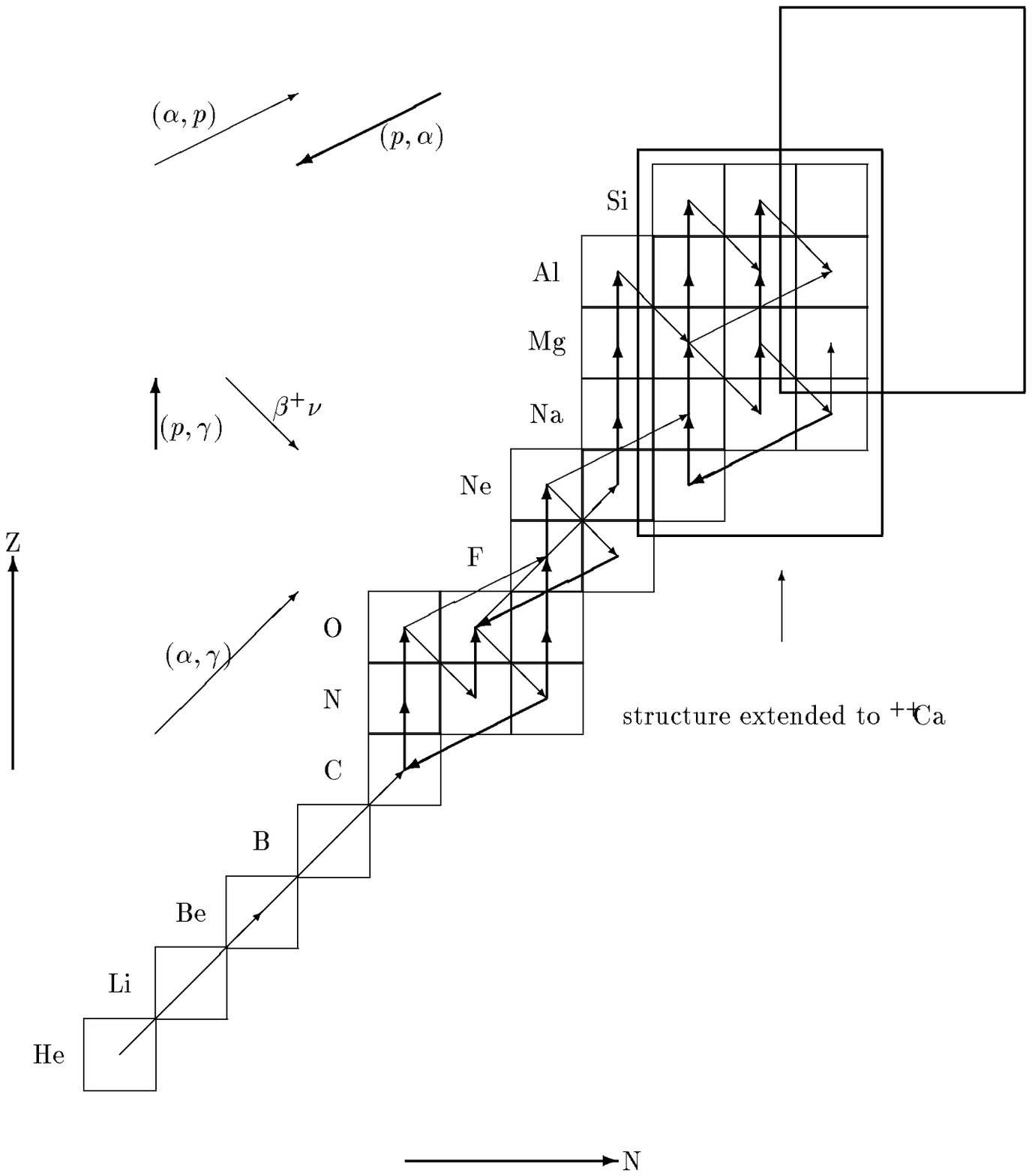,width=11cm,angle=0}}
\vspace{-4cm}
\caption{
The hot CNO-cycle incorporates three proton captures,
($^{12}$C, $^{13}$N, $^{14}$N), two $\beta^+$-decays ($^{14,15}$O) and
one (p$,\alpha)$-reaction ($^{15}$N). In a steady flow of reactions
the long beta-decay half-lives are responsible for high abundances of 
$^{15,14}$N (from $^{15,14}$O decay) in nova ejecta. 
From Ne to Ca, cycles similar to the hot CNO exist, based always
on alpha-nuclei like $^{20}$Ne, $^{24}$Mg etc. The exception is the 
(not completed) cycle based on $^{16}$O, due to 
$^{18}$F(p,$\alpha)^{15}$O, which provides a reaction path
back into the hot CNO-cycle. Thus, in order to proceed from C to heavier nuclei,
alpha-induced CNO break-outs are required. The shown flow pattern, which
includes alpha-induced reactions, applies for temperatures in the range 
4-8$\times 10^8$~K. Smaller temperatures permit already processing of 
pre-existing Ne via hot CNO-type cycles. This leads to the typical nova
abundance pattern with $^{15,14}$N enhancements, combined with specific
isotopes up to about Si or even Ca. If the white dwarf is an ONeMg rather
than CO white dwarf, containing Ne and Mg in its initial composition resulting 
from carbon 
rather than helium burning, the latter feature is specifically recognizable.}
\end{figure}

\begin{figure*}[t]
\psfig{file=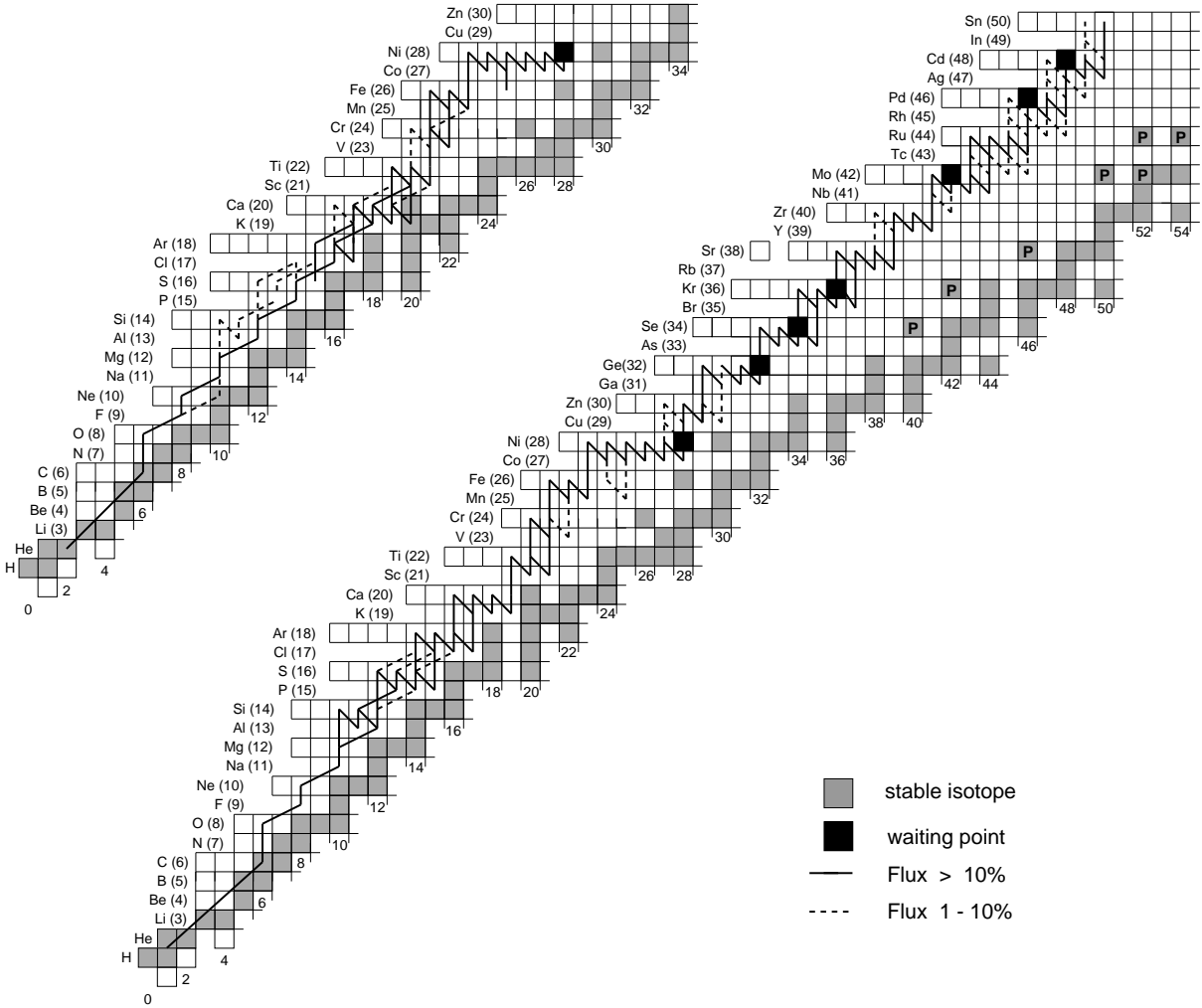,width=16cm,angle=0}
\caption{
rp and $\alpha$p-process flows up to and beyond Ni. The reaction
flows shown in the nuclear chart are integrated reaction fluxes from a 
time dependent network calculation (Wiescher \& Schatz 2000), (a) during the
initial burst and thermal runaway phase of about 10s (top left), (b) after the 
onset of the cooling phase when the proton capture on $^{56}$Ni is not blocked 
anymore by photodisintegrations (extending for about 200s, bottom right). 
Waiting points above 
$^{56}$Ni are represented by filled squares, stable nuclei by hatched squares,
light p-process nuclei below A=100 are indicated by a $P$.}
\label{rp1}
\end{figure*}
In essentially all nuclei below Ca, a proton capture reaction on the
nucleus ($Z_{even}$--1,$N$=$Z_{even}$) produces the compound nucleus
above the alpha-particle threshold and permits a (p,$\alpha)$ reaction. 
This is typically not the case for ($Z_{even}$--1,$N$=$Z_{even}$--1) due to 
the smaller proton separation energy and leads to hot CNO-type cycles
above Ne (Thielemann et al.\ 1994, see Fig.\ 11). There is one exception, 
$Z_{even}$=10, 
where the reaction $^{18}$F(p,$\alpha)$ is possible, avoiding $^{19}$F
and a possible leak via $^{19}$F(p,$\gamma)$ into the NeNaMg-cycle.
This has the effect that only alpha induced reactions like $^{15}$O($\alpha,
\gamma)$ can aid a break-out from the hot CNO-cycle
to heavier nuclei beyond Ne (Wiescher et al.\ 1999).
Break-out reactions from the hot CNO-type cycles above Ne proceed typically via
proton captures on the nucleus ($Z_{even}$,$N$=$Z_{even}$-1)
and permits a faster build-up of heavier nuclei
(Thielemann et al.\ 1994, Rembges et al.\ 1997). They occur
at temperatures of about $3\times 10^8$K, while the alpha-induced break-out 
from the hot CNO-cycle itself is delayed to about $4\times 10^8$K.

\begin{figure*}[t]
\centerline{\psfig{file=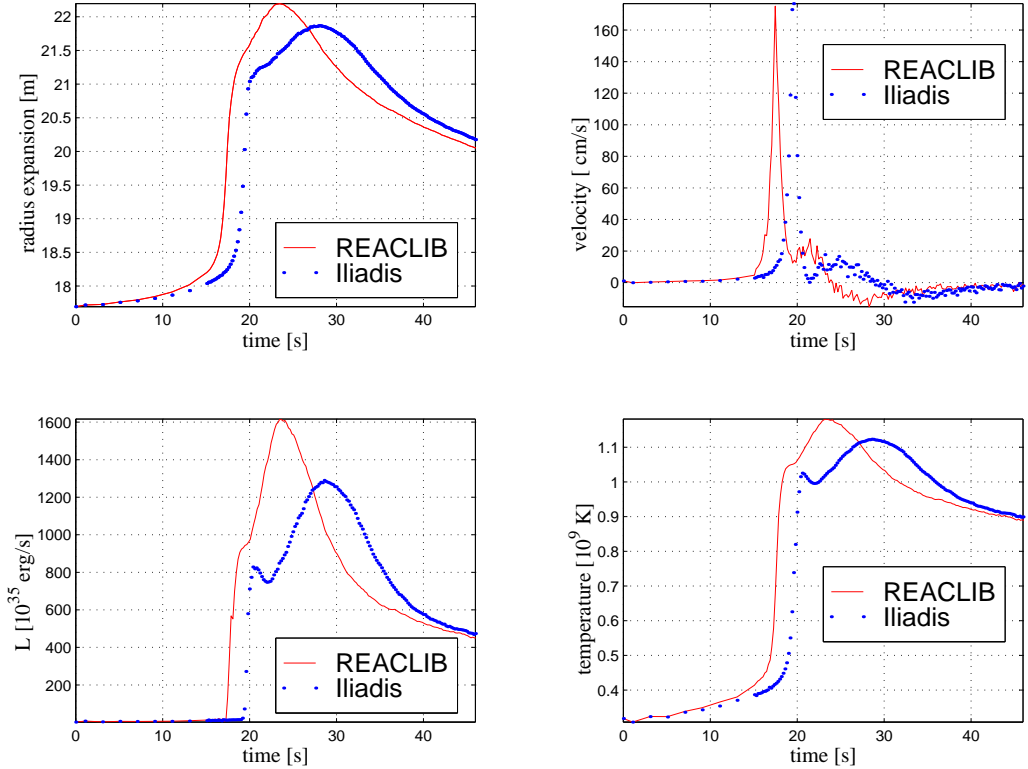,width=16cm,angle=0}}
\caption{
Comparison of burst profiles with different proton capture rates for
the break-out reactions on $^{27}$Si, $^{31}$S, $^{35}$Ar, and $^{38}$Ca
in a self-consistent X-ray burst model (Rembges 1999, Rauscher et al.\ 2000). 
Due to different reaction rates, matter beyond Ne is burned on different
timescales, causing a different pre-expansion before the maximum temperature 
is attained during alpha-induced reaction sequences.}
\label{rp2}
\end{figure*}

Explosive H-burning in novae has been discussed in many recent articles
(Jos\'e et al.\ 1999, Starrfield 1999, Starrfield et al.\ 2000, Coc et al.\ 2000,
Jos\'e et al.\ 2001). Its processing seems limited
due to maximum temperatures of $\sim$3$\times 10^8$K. Apparently,
only in X-ray bursts temperatures larger than 4$\times 10^8$K are possible.
This is important, because it leads to a quite different outcome of 
hot/explosive H-burning. The major nucleosynthesis yields for novae are 
therefore related to hot CNO-products like $^{15}$N and specific nuclei between
Ne and Si/S, which are based on processing of pre-existing Ne (more details
are discussed together with Fig.\ 11).

\subsection{Type I X-ray bursts}

In type I X-ray bursts (Taam 1985, Lewin et al.\ 1993, Taam et al.\ 1996, 
Schatz H. et al.\ 1998, 2001, Wiescher \& Schatz 2000) hydrogen, and 
subsequently also helium, burn 
explosively in a thermonuclear runaway. Once the hot CNO-cycle, and hot CNO-type
cycles beyond Ne, have generated sufficient amounts of energy in order to
surpass temperatures of $\sim$4$\times 10^8$K, alpha-induced reactions
lead to a break-out from the hot CNO-cycle. This provides new fuel for
reactions beyond Ne, leading to a further increase in temperature.

In the next stage of the ignition process also He is burned
via the 3$\alpha$-reaction $^{4}$He($2\alpha,\gamma)^{12}$C, filling the
CNO reservoir, and the 
$\alpha$p-process (a sequence of ($\alpha,$p) and (p,$\gamma$) reactions,
Wallace \& Woosley 1981) sets in. It produces nuclei up to Ca and 
provides seed nuclei for hydrogen burning via the rp-process
(proton captures and beta-decays).
Processing of the $\alpha$p-process and rp-process up to and beyond $^{56}$Ni 
is shown in  Fig.\ \ref{rp1} (Wiescher \& Schatz 2000).
Certain nuclei play the role of long waiting points in 
the reaction flux, where long beta-decay half-lives
dominate the flow, either competing with slow ($\alpha$,p) reactions
or negligible (p,$\gamma)$ reactions, because they are inhibited by inverse
photodisintegrations for the given temperatures. Such nuclei were identified 
as $^{25}$Si ($\tau_{1/2}$=0.22s), $^{29}$S (0.187s),
$^{34}$Ar (0.844s), $^{38}$Ca (0.439s). 
The bottle neck at $^{56}$Ni
can only be bridged for minimum temperatures around $10^9$K (in order to
overcome the Coulomb barrier for proton capture) and maximum temperatures below
$2\times 10^9$K (in order to avoid photodisintegrations), combined with
high densities exceeding $10^6$g~cm$^{-3}$ which support the capture process
(Schatz et al.\ 1998, Rehm et al.\ 1998).
If this bottle neck can be overcome, other waiting points like $^{64}$Ge
(64s), $^{68}$Se (96s), $^{74}$Kr (17s) seem to be hard to pass. However,
partially temperature dependent half-lives (due to excited state population),
or mostly 2p-capture reactions via an intermediate proton-unstable nucleus
(introduced in G\"orres et al.\ 1995 and
applied in Schatz et al.\ 1998) can help. The final endpoint of the rp-process
was found recently by Schatz et al.\ (2001), who determined a closed reaction
cycle in the Sn-Sb-Te region, due to increasing alpha-instability of heavy 
proton-rich nuclei.

The initial break-out reactions from hot CNO-type cycles and the
hold-ups at waiting points introduce a time structure in energy generation. 
One of the essential questions is whether they can/will show up in 
the X-ray light-curves of bursts. The other question is whether 
individual proton-capture reactions matter, because at peak temperatures 
(p,$\gamma)$-$(\gamma,$p) equilibria are attained, determined only by nuclear
mas\-ses of the nuclei involved. This leads to a chemical equilibrium
distribution for proton-capture reactions along isotonic lines and a steady 
flow of beta-decays between these equilibrium abundance maxima.
Therefore, Iliadis et al.\ (1999) claimed that individual (p,$\gamma$)-reactions
do not influence burst properties, based on nucleosynthesis calculations
utilizing an X-ray burst temperature
profile provided by Rembges (1999). This paper showed (as should 
be expected from the final equilibrium conditions) that, with a given temporal 
temperature profile, the resulting composition does not depend on individual 
reactions. The important question, however, is whether the feedback from
hydrodynamics, due to the changed energy input in the early ignition stages
during the working of 
break-out reactions, leads to different temperature profiles
and thus different final abundances. 

The results of a self-consistent calculation can be seen in Fig.\ \ref{rp2}
from (Rembges 1999, Rauscher et al.\ 2000), 
which shows the radius of the burning shell,
the velocity, the luminosity, and the temperature. The difference between
calculations with the old REACLIB rates (Thielemann et al.\ 1987,
Cowan et al.\ 1991) and the ones
with updated cross sections for the break-out points $^{27}$Si, $^{31}$S,
$^{35}$Ar, and $^{38}$Ca emerges at the expected temperatures of
$(3-4)\times 10^8$K, when these break-out reactions occur. At that point
the nuclei beyond Ne will burn towards Ni/Fe. This energy input causes
the temperature increase which will at first permit the hot CNO break-out via
alpha-induced reactions, followed by the triple-alpha link to $^{12}$C.
This leads to the burning of $^4$He and determines the temperature peak of the
rp-process with a chemical equilibrium for proton-capture reactions.
However, as the initial break-out phase differs, different pre-expansions can
occur, causing different densities and also different peak temperatures.

This effect was recently underlined in further self-consistent burst 
calculations (Fisker et al.\ 2001b, Wiescher et al.\ 2002), when REACLIB rates 
in the mass range $A$=44--63 were
replaced by cross sections based on resonance properties
from shell model calculations (Fisker et al.\ 2001a). These effects are even more
drastic, again due to the early burning phase when matter beyond Ne
burns up to Fe, before the alpha-captures begin. This shows that
a more precise determination of specific reaction rates is important, when
self-consistent network plus hydrodynamics calculations are performed.
The peak temperatures and densities attained in X-ray burst calculations
depend on this cross sections input.
They depend, however, also on a correct modeling of the deeper neutron star
crust layers, which consist of ashes of previous bursts. Electron captures
due to higher densities in deeper layers and ignition of previously unburned
ashes can lead to higher temperatures at the inner boundaries of the unburned
H/He-layer and even cause so-called "superbursts" (Cummings \& Bildsten 2001).
Thus, it is paramount not just to model the H/He-layers with inner boundary
conditions (Rembges 1999, Rauscher et al.\ 2000, Fisker et al.\ 2001b, Wiescher
et al.\ 2002), but to model consistently the deeper layers as well.

If only a small percentage of the synthesized matter in X-ray bursts escapes 
the strong
gravitational field of the neutron star, some proton-rich stable nuclei
(p-process nuclei) below $A$=100 (indicated
as P in Fig.\ \ref{rp1}) could be explained in the solar system abundances.
The more massive p-process isotopes (contributing only 1\% or less to their
element abundances) are probably due to $(\gamma,$n) and $(\gamma,\alpha)$
"spallation" reactions of pre-existing stable nuclei during explosive burning
in those mass zones of supernovae which attain temperatures of about
$2\times 10^9$K (Woosley \& Howard 1978, Costa et al.\ 2000, Rauscher et
al.\ 2001b, Rauscher et al.\ 2002).

\section{Heavy Elements: r- and s-Process}

\begin{figure*}[t]
\vspace*{1.5cm}
\figbox*{}{}{\psfig{file=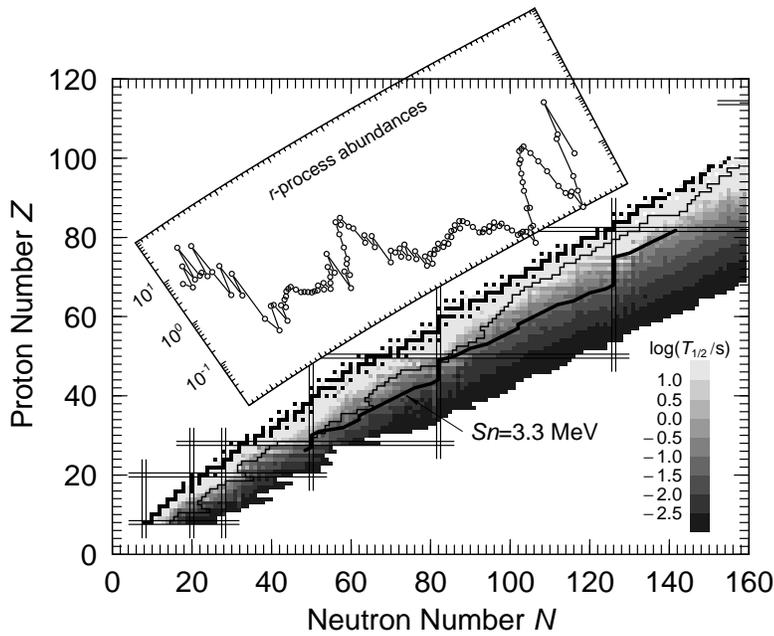,angle=-90,width=10cm}}
\vspace*{-6.8cm}
\caption{Some features of nuclei in the ($N,Z$)-chart of isotopes. Stable nuclei
are indicated by black filled squares. The thin solid line represents the
present limit of experimentally known nuclear masses. The magic numbers
are shown as double bars. The thick solid line is the contour line of
constant neutron separation energy $S_n$=3.3~MeV. It relates nuclear 
properties to astrophysical abundances of the r-process.
It can be recognized that the abundances are proportional to the
$\beta^-$-decay half-lives [indicated by grey shades in $\log_{10}(\tau_{1/2})$]
along $S_n$ contour lines.}
\end{figure*}

\begin{figure*}[t]
\vspace*{1cm}
\figbox*{}{}{\psfig{file=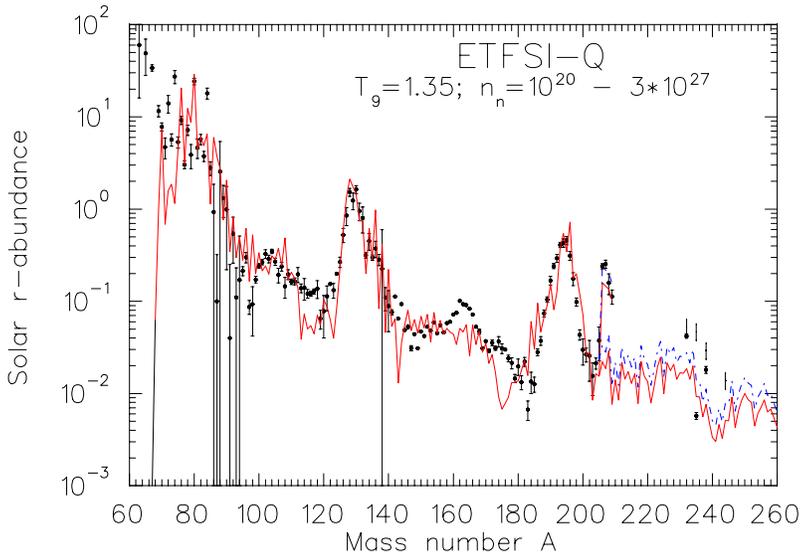,angle=90,width=10.5cm}}
\vspace*{-0.9cm} \caption{Fits to solar r-process abundances,
obtained with two different smooth superposition of 17 equidistant 
$S_n(n_n,T)$ components from 1 to 4 MeV (solid and dashed lines).
The ETFSI-Q mass model (Pearson et al.\ 1996) was applied,
which introduces a phenomenological quenching of shell effects.
The quenching of the $N=82$ shell gap avoids a large abundance
trough below the $A$=130 peak. These results also
show a good fit to the r-process
Pb and Bi contributions after following the decay chains of unstable heavier
nuclei (indicated by two sets - dashed lines - of abundances for A$>$205).
\label{ueberlagerung}}
\end{figure*}

\subsection{Neutron capture processes}

It is long understood that the existence of the heavy elements in nature
is due to neutron capture (Suess \& Urey 1956) and that (at least) two types of 
processes must be responsible (Burbidge et al.\ 1957, Cameron 1957):
1. A process with small neutron densities, experiencing long neutron capture
timescales in comparison to $\beta$-decays ($\tau_\beta < \tau_{n,\gamma}$,
slow neutron capture or the s-process), causing abundance peaks in the flow 
path at nuclei with small neutron capture cross sections. These are stable 
nuclei with closed shells, i.e. magic neutron numbers (K\"appeler et al.\ 1989). 
2. A process with high neutron densities and temperatures, experiencing rapid 
neutron captures and the reverse photodisintegrations
with $\tau_{n,\gamma},\tau_{\gamma,n} < \tau_\beta$. This leads to chemical
equilibria for neutron captures along isotopic chains and the
rapid neutron-capture process (r-process) produces  
highly unstable nuclei with short half-lives (Kratz et al.\ 1993,
K\"appeler et al.\ 1998, Pfeiffer et al.\ 2001), permitting also the formation 
of the heaviest elements in nature like Th, U, and Pu. 
The r-process involves nuclei near the neutron drip-line,
where the last neutron is not bound and thus the neutron binding or neutron
separation energy is zero or negative. Far from stability,
neutron shell closures are encountered for smaller mass
numbers than in the valley of stability. Therefore, 
a steady flow of beta-decays between the maxima of isotopic equilibrium chains
causes r-process peaks due to long $\beta$-decay half-lives. 
These are encountered closest to stability at neutron-magic nuclei.
Therefore, the r-process
abundance peaks are shifted in comparison to the s-process peaks 
(which occur for neutron shell closures at the stability line).

This behavior is indicated in Fig.\ 14. The r-process runs along equilibrium 
maxima at constant neutron separation energy. One such line
with $S_n$=3.3MeV is shown in the nuclear chart. The peaks are given
by the longest beta-decay half-lives (closest to stability at shell closures).
After decay, this results in an abundance distribution as a function of mass
number $A$ as shown in the inserted graph.
The s-process maxima occur for the smallest neutron capture cross sections at 
closed shells (magic numbers) along the line of stability.
Therefore, the r-process peaks are shifted to lower
mass numbers. The decomposition of solar abundances of heavy nuclei into
s- and r-process components has been shown in Fig.\ 2 and presented with a
basic discussion of the s-process and its sites. Here we want to concentrate
on the r-process.

Besides this basic understanding, the history of r-process research has
been quite diverse in suggested scenarios (for reviews see 
Cowan et al.\ 1991, Wallerstein et al.\ 1997, K\"appe\-ler et al.\ 1998, Pfeiffer et 
al.~2001). If starting with a seed distribution somewhere around
A=50-80, before rapid neutron-capture sets in, the operation of an r-process 
requires 10 to 150 neutrons per seed nucleus to form all heavier r-nuclei.
The question is which kind of environment can provide such a supply of 
neutrons which need to be utilized before decaying with a 10 min half-life.
The logical conclusion is that only explosive environments, producing
or releasing these neutrons suddenly, can account for such conditions.
Two astrophysical
settings are suggested most frequently: (i) Type II supernovae with 
postulated high-entropy ejecta (Woosley et al.\ 1994, Takahashi et al.\ 1994,
Qian \& Woosley 1996, Meyer et al.\ 1998, Otsuki et al.\ 2000, 
Freiburghaus et al.\ 1999a, Nagataki \& Kohri 2001, Wanajo et al.\ 2001,
Thompson et al.\ 2001)
and (ii) neutron star mergers or similar events (like axial jets in 
supernova explosions) which eject neutron star matter with low-entropies 
(Lattimer et al.\ 1977, Meyer 1989, Eichler et al.\ 1989, Freiburghaus et al
1999b, Rosswog et al.\ 1999, 2000, 2001, Cameron 2001ab).
These two sites are representative for two (high or low) entropy options.

\subsection{Working of the r-process and solar r-abundances} 

First, we want to discuss here some general features.
The high neutron density and temperature 
environments, leading to an r-process, result in 
a chemical equilibrium for neutron captures and the reverse 
photodisintegrations ($n,\gamma)\rightleftharpoons(\gamma,n)$ 
within isotopic chains for each element and cause abundance
maxima at a specific neutron separation energy. 
Thus, the combination of a neutron density $n_n$ and temperature 
$T$ determines the r-process path defined by 
a unique neutron separation energy $S_n$. 
Therefore a choice of either ($n_n,T$) or $S_n$ for an r-process
is equivalent.
During an r-process event exotic nuclei with
neutron separation energies of 4 MeV and less are important, up to $S_n$=0, 
i.e. the neutron drip-line. This underlines that the understanding of
nuclear physics far from stability is a key ingredient (Kratz et al.\ 1998).

Traditionally, before having knowledge of a particular r-process model (site),
attempts were made to fit solar r-process abundances with a choice of
neutron number densities $n_n$, temperatures $T$ and exposure timescales $\tau$.
Our r-process parameter studies showed that the entire isotopic abundance 
pattern cannot be reproduced by assuming a unique r-process path,
characterized by one neutron separation energy
(Thielemann et al.\ 1993, Kratz et al.\ 1993). Instead, even in a single 
astrophysical event it requires a
superposition of a multitude of r-components (a minimum of three) with different
neutron densities or equivalently different S$_n$'s, related to different
r-process paths and time scales (Chen et al.\ 1995, Goriely \& Arnould 1996, 
Cowan et al.\ 1999, Freiburghaus et al.\ 1999a).
In Fig.\ 15 we show global abundance distributions from a superpostion of
sixteen n$_n, T, \tau$ components, utilizing the ETFSI-Q (Pearson et al.\ 1996)
nuclear mass model.  The successful reproduction
of the position and relative height of the peaks, as well as the
remaining deficiencies, have been interpreted as signatures of nuclear 
structure near the neutron drip-line (Thielemann et al.\ 1994, Chen et al.\ 1995,
Kratz et al.\ 1998, Kratz et al.\ 2000, Pfeiffer et al.\ 2001). 

A more realistic astrophysical setting is given by an explosion (leading
to initial entropies) and adiabatic expansion (Hoffman et al.\ 1996, 1997).
In a superposition of high-entropy 
environments, the high entropies (up to 400 $k_b$/nucle\-on, responsible for the 
mass region A$>$110) reproduce nicely the r-abundance pattern. Some exceptions
in the mass region 110--120 are related - as in the site-independent 
studies with $n_n, T, \tau$ - to nuclear structure effects at the shell 
closures N=82. Thus, both approaches can have a nearly one-to-one mapping 
between their resulting abundance features (Freiburghaus et al.\ 1999a). 
Discrepancies occur in the
mass region 80--110, for which lower entropies are responsible. 
The reason is that essentially no neutrons are left after the freeze-out of
charged-particle reactions with such entropies for $Y_e=0.45$. 
Then the abundance pattern is not reflecting  
neutron-separation energies. Possibly more neutron-rich environments with
smaller $Y_e$'s could improve this feature, which then do not require
(such) high entropies (Meyer 1989, Freiburghaus et al.\ 1999a).

\begin{figure*}[t]
\centerline{\psfig{file=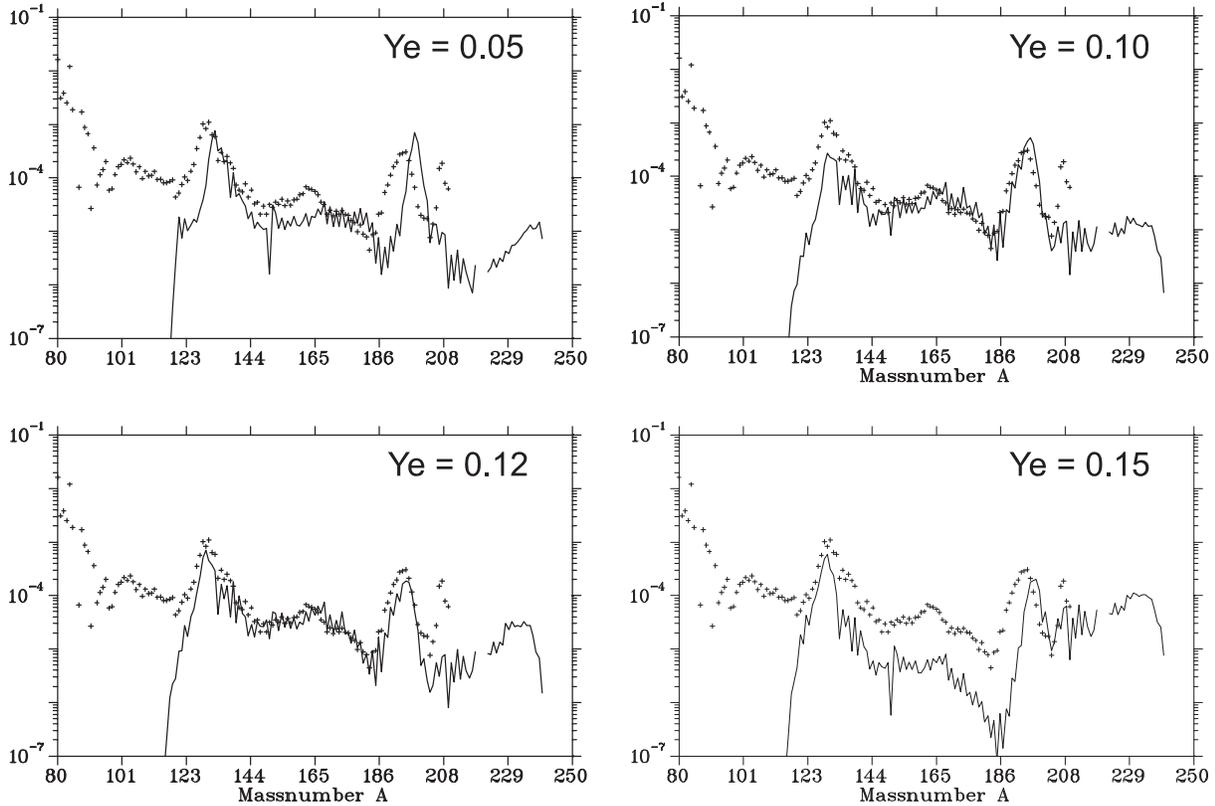,angle=0,width=16cm}}
\caption{Calculated r-process distributions for neutron star ejecta with
different initial $Y_e$'s. In general one obtains useful contributions for
$0.08<Y_e<0.15$. A further discussion is given in the text.
$Y_e$ determines the total neutron/seed ratio, which is an indication
of the strength of the r-process. It affects also the combination
of $n_n$ and $T$, i.e.\ the r-process path, and therefore the position
of peaks. Finally, fission cycling is responsible for the drop of abundances
below A=130, but only an improved incorporation of fission barriers and 
yields will provide the correct abundance distribution in this mass range.}
\label{fig4}
\end{figure*}

\subsection{Astrophysical sites of the r-process}

We have shown in section 6.2 that with the known nuclear properties a 
site-independent approach, based on a superposition of neutron densities
or entropies, can reproduce the solar system
r-process abundance pattern (at least beyond Ba). Low metallicity stellar
observations show that one type of astrophysical event is responsible for this
main r-process component beyond A=130, while another possibly weaker source
fills in the lower mass numbers (see section 7). Now we need to address the 
question of the related astrophysical sites.
The necessary conditions turned out to occur in either high
entropy environments, which can be moderately neutron-rich, or low entropy
environments, which have to be very neutron-rich. The sites which obviously
come into mind are SNe II and neutron-star (merger) ejecta.
However, for a historical review of all sites suggested so far see Cowan et al.\
et al.\ (1991) and Wallerstein et al.\ (1997), including cosmological, i.e. big 
bang, sources (Orito et al.\ 1997).

If SNe II are also responsible for the solar r-process abundances, given the
galactic occurance frequency, they would need to eject about $10^{-5}$
M$_\odot$ of r-process elements per event (if all SNe II contribute equally).
The scenario is based on the so-called ``neutrino wind'', i.e.
a wind of matter from the neutron star surface
(within seconds after a successful supernova explosion) is driven via neutrinos
streaming out from the still hot neutron star (Woosley et al.\ 1994, Takahashi
et al.\ 1994, Hoffman et al.\ 1996, 1997, Qian \& Woosley 1996,
Meyer et al.\ 1998, Otsuki et al.\ 2000).

This high entropy neutrino wind is expected to lead to a superposition of
ejecta with varying entropies.
The r-process by neutrino wind
ejecta of SNe II faces two difficulties: (i) whether the required high entropies
for reproducing heavy r-process nuclei can really be
attained in supernova explosions has still to be verified
(Rampp \& Janka 2000, Mezzacappa et al.\ 2001, Liebend\"orfer et al.\ 2001ab,
Nagataki \& Kohri 2001, Thompson et al.\ 2001), (ii)
the mass region 80--110 experiences difficulties to be reproduced adequately
(Freiburghaus et al.\ 1999a, Wanajo et al.\ 2001).
It has to be seen whether the inclusion of non-standard
neutrino properties (McLaughlin et al.\ 1999) can cure both difficulties or
lower $Y_e$ zones can be ejected from SNe II, as recently claimed (Sumiyoshi
et al.\ 2001) from assumed prompt, but probably unrealistic, explosion 
calculations lacking a proper neutrino transport (see the discussion of 
recent literature in section 3).

Fig.\ 4 showed the abundance evolution in the innermost zone of a
SNe II simulation (Hauser 2002), based on Lieben\-d\"orfer et al.\ (2001a) but 
with (artificially) reduced neutrino opacities by 60\%. This led to larger
neutrino luminosities, permitting a successful delayed explosion. 
Entropies per bary\-on of up to 50 are attained, which is too small for a
strong r-process (Freiburghaus et al.\ 1999a). However,
we see a freeze-out with remaining neutrons, leading to the onset
of a (weak?) r-process, indicated by the rise of Ge beyond Ni (Ge is the
upper limit of the nuclear network employed in that calculation). This result 
seems possible
as a combination of high entropies and a low $Y_e$=0.39, but needs further
analysis. In any case, we have only indications for a weak r-process.
Another supernova related site responsible for the ``weak'' r-process component
(i.e. nuclei with $A$$<$130) could also
be related to explosive C or He-burning in the outer ejected shells
(Thielemann et al.\ 1979, Wheeler et al.\ 1998, Truran et al.\ 2001,
Meyer et al.\ 2000, Rauscher et al.\ 2001a, Rauscher et al.\ 2002).

An alternative site for the main component are neutron-star ejecta, like e.g.\ 
in neutron star mergers. The
binary system, consisting of two neutron stars, looses energy 
and angular momentum through the emission of gravitational 
waves and merges finally. Such systems are known to exist; four NS-NS 
binaries have been detected by now (Thorsett 1996, Lorimer 2001). 
The measured orbital decay gave the first evidence for the existence of 
gravitational radiation (Taylor 1994) and indicates timescales of the
order of $10^8$y or less (dependent on the excentricity of the system). 
The rate of NS mergers has been estimated to be of the order
$10^{-6}-10^{-4}$y$^{-1}$ per galaxy (Eichler et al.\ 1989, Narayan et al.\ 1992);
more recent estimates (van den Heuvel \& Lorimer 1996, Ka\-lo\-ge\-ra et al.\ 2001) 
tend towards the center of this range ($8\cdot10^{-6}$y$^{-1}$ per galaxy). 
A merger of two NS can lead to the ejection of neutron-rich material 
(Davies et al.\ 1994, Janka \& Ruffert 1996, Baumgarte et al.\ 1997,
Ruffert \& Janka 1998, Rosswog et al.\ 1999, 2000, Ruffert \& Janka 2001)
of the order of $10^{-2}$M$_\odot$ in Newtonian calculations, and could be a 
promising site for the production of r-process 
elements. It is even possible that such mergers account for {\it all} heavy 
r-process matter in the Galaxy 
(Lattimer et al.\ 1977, Eichler et al.\ 1989, Freiburghaus et al.\ 1999b,
Rosswog et al.\ 2001). 
The decompression of cold neutron-star matter has been studied 
(Lattimer et al.\ 1977, Meyer 1989),
however, a hydrodynamic calculation coupled with 
a complete r-process calculation has not been undertaken, yet. 

Fig.\ 16 shows the composition of ejecta from a NS merger, dependent on
the assumed (not self-consistently modeled) $Y_e$ of the ejecta
(Freiburghaus et al.\ 1999b, Rosswog et al.\ 2001). It is seen 
that the large amount of free neutrons (up to n$_n$$\simeq$10$^{32}$ cm$^{-3}$)
available in such a scenario leads to the build-up of the heaviest elements
and also to fission cycling within very short timescales, while the flow from 
the Fe-group to heavier elements "dries up". This leads to a 
composition void of abundances below the A$\simeq$130 peak, which is, 
however, dependent on detailed fission yield predictions (Panov et al.\ 2001).
If further observations support such a behavior of Z$<$50 elements in very low 
metallicity stars (see section 7), it would provide strong support for that 
type of r-process site, 
but would definitely require an additional weak astrophysical source which 
produces the bulk of the lighter r-abundances up to A$\simeq$125.
Additional important evidence for the source composition of r-process
elements could come from the actinide cosmic ray composition
(Westphal et al.\ 1998, 2001).

At present, these
suggested r-process sources, supernovae and neutron star mergers (jets),
did not yet prove to be "the" main-component r-process source without reasonable
doubt. Self-consistent core collapse supernovae do not give explosions
(Ramp \& Janka 2000, Mezzacappa et al.\ 2001, 
Lieben\-d\"orfer et al.\ 2001a, 2002), yet, but parameter studies with
neutrino opacities permit to "fit" the correct explosion behavior
(Hauser et al.\ 2002). Thus, there is no way to predict
whether the required entropies for an r-process can be obtained
(Wanajo et al.\ 2001, Thompson et al.\ 2001). Hypernovae (massive stars which end with a central
stellar mass black hole rather than a neutron star after supernova explosions)
could do so, but their full magneto-hydrodynamic understanding is also not
revealed, yet (MacFadyen \& Woosley 1999, Cameron 2001ab, MacFadyen et al.\ 
2001).  Neutron star merger 
calculations give the correct mass ejection (Rosswog et al.\ 1999, 2000), but 
until now only for non-relativistic calculations. First relativistic merger 
calculation seem to eject smaller amounts of matter (Oechslin et al.\ 2002).
The abundance predictions for neutron star
mergers look excellent (Frei\-burghaus et al.\ 1999b), but still take $Y_e$ as a 
free
parameter rather than treating weak interactions and neutrino transport
self-consistently.

\section{Observational Constraints and Galactic Evolution}

\subsection{Observations of individual sources}

Explosive nucleosynthesis yields leave fingerprints in observations of 
individual stellar events as well as their remnants via spectra, lightcurves,
X-rays and radioactivities/decay gam\-ma-rays (see e.g.\ the discussions in
Thielemann et al.\ 1996, 2000, Iwamoto et al.\ 1999). 

$^{56,57}$Ni
and $^{44}$Ti abundances, and possibly stable Ni/Fe ratios in SNe II, give
insight into the details of the explosion mechanism with respect to
the mass cut between the neutron star and the SN ejecta, the total energy
of the explosion, the entropy and the $Y_e$ in the innermost ejecta
(see section 3). The intermediate mass elements Si-Ca
provide information about the explosion energy and the stellar structure
of the progenitor star, while elements like O, Ne and Mg are essentially
determined by the stellar progenitor evolution.
Therefore, only the correct reproduction of Fe-group abundances are a direct
test whether one understands the explosion mechanism (Thielemann et al.\ 2001).

In SNe Ia, the $^{56}$Ni production and the Si-Ca/Fe ratio are related
to the total explosion energy and burning front speed in layers of the 
exploding white dwarf.
Constraints on the ignition density and burning front speed in the central
regions (as discussed in section 4) are reflected e.g.\ in minor isotopic
abundances like $^{50}$Ti and $^{54}$Cr, where direct supernova or remnant
observations cannot be used. Here one can only make use of global abundance
constraints from galactic evolution, which therefore permit only 
statements about an "average" SN Ia. The application of present day
electron capture rates makes it hard to account for our solar system $^{48}$Ca
(Brachwitz et al.\ 2002).
On the other hand, observations of varying stable Ni/Fe ratios during the
evolution of a supernova (by spectral means) related to $^{54}$Fe and
$^{58}$Ni, might give clues to the metallicity of the exploding white dwarf
(H\"oflich et al.\ 1998).

Observations of nova ejecta support the features discussed in section 5
(Truran 1998), while the spectral resolution of new X-ray observatories
can help to put constraints on X-ray burst models.
So far, there have been no direct observations of (rare) r-process elements in
individual explosions, but the cosmic-ray source composition of actinides
(Westphal et al.\ 1998, 2001) might provide specific clues.

\subsection{Galactic evolution}

Opposite to individual source observations, 
galactic evolution can serve as a global test for all contributing stellar 
yields,
i.e. especially intermediate and low mass stars through planetary nebula
ejection, mass loss from massive stars, as well as the ejecta of SNe II and
SNe Ia (for details see e.g.\
Tsujimoto et al.\ 1995, Thomas et al.\ 1998, Matteucci et al.\ 1999, Chiappini et 
al. 1999). The composition of interstellar gas in galaxies as well as
clusters of galaxies can serve the same purpose (Finoguenov et al.\ 2002).
Here we want to focus on specific observational
clues to SNe Ia and SNe II nucleosynthesis and indicate how very low metallicity
stars might witness individual rather than only integrated SNe II yields.

Stars (with understood exceptions) do not change the surface composition during
their
evolution. Thus, surface abundances reflect the interstellar medium (ISM, out of
which the stars formed) at the time of their formation. Therefore, observations
of the surface composition (via spectra) of stars over a variety of ages 
and metallicities give a clue to gas abundances throughout the
evolution of our Galaxy. 

Very old (low metallicity) stars witness very early galactic evolution and
thus fast evolving massive
stars, i.e.~SNe II nucleosynthesis, seen in an enhancement of the alpha 
elements (O through Ca) in comparison to Fe for [Fe/H]=${\rm log_{10}}$
[(Fe/H)/(Fe/H)$_\odot ]<$$-$1 (Gratton \& Sneden 1991, Nissen 
et al.\ 1994, Argast et al.\ 2000, 2001). The higher ratio of Fe-group 
elements to Si-Ca in SNe Ia, appearing later in galactic evolution, has to 
compensate for these overabundances in 
SNe II in order to obtain solar abundance ratios for the combined 
nucleosynthesis products at solar metallicity [Fe/H]=0.

The Fe-group ejecta of both types of supernovae have to differ because Ti/Fe and
Mn/Fe are found above, or respectively below, solar ratios in SNe II, requiring
the opposite behavior of SNe Ia (Iwamoto et al.\ 1999). At very low metallicities
Cr, Mn, and Co show peculiar features, possibly related to the progenitor mass
dependence of SNe II (Nakamura et al.\ 1999).

Recent promising trends in galactic evolution
modeling might provide constraints on individual supernova models
rather than only global properties of SNe II and SNe Ia. The reason for
this possibility is the fact that there is no instantaneous mixing of ejecta
with the interstellar medium, and therefore early phases of galactic evolution can
present a connection between low metallicity star observations and a 
single supernova event. On average, each supernova pollutes a volume of the 
interstellar medium containing $\approx$ $(3-5)\times 10^4$M$_\odot$. 
(for references see e.g.\ Argast et al.\ 2000). Each volume of the 
interstellar medium containing $\approx$ $3\times 10^4$M$_\odot$ needs to be
enriched by $\approx$ $10^3$ SNe in order to obtain solar metallicities.

After a supernova polluted such a previously pristine 
environment mass, it results in values for [x/Fe] and
[Fe/H] in the remnant, x standing for any element. Variations in SN II 
properties as a function of
progenitor mass (and metallicity) should lead to a scatter in [x/Fe] for 
the same [Fe/H]. 
The amount of the polluted volume depends on the explosion energy, and if
there is a strong variation of explosion energies with progenitor mass
this could affect the relation between [x/Fe] and the metallicity [Fe/H]
(Nakamura et al.\ 1999).
The scatter in [x/Fe] expected for the same [Fe/H] is observed up to 
metallicities of [Fe/H]=-2, where it vanishes because overlapping contributions
from many SNe II
behave like a well mixed medium (Argast et al.\ 2000, 2001).

\begin{figure*}[t]
\centerline{\psfig{file=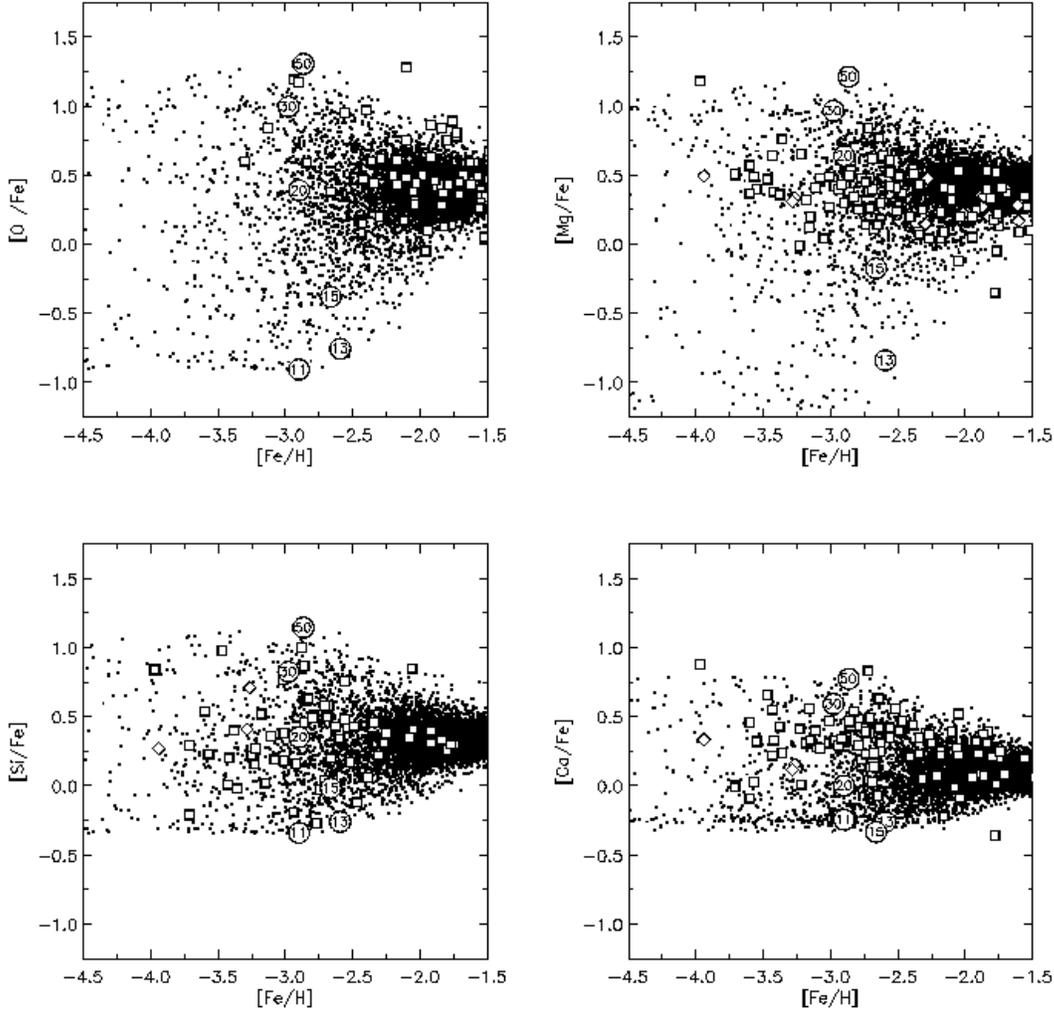,angle=0,width=14.3cm}}
\caption{Comparison of low metallicity observations with SN II yields and
a galactic evolution model.  Three
types of symbols are given, circles, squares and dots. The {\it squares show
observations} of low metallicity stars with a scatter at very low 
metallicities.
The {\it open circles} indicate the x/Fe and Fe/H ratios in a volume of
$5\times 10^4$M$_\odot$ of an initially pristine (big bang) composition,
polluted by {\it single supernovae} with our ejecta compositions. 
The numbers inside the
circles refer to the progenitor mass of a supernova whose ejecta composition
was mixed with $5\times 10^4$M$_\odot$ of prestine ISM.
The (galactic evolution) {\it model stars are indicated by dots} (see the text 
for a detailed discussion).}
\label{fig:golferbig}
\end{figure*}

Some results from such an approach (Argast et al.\ 2000, 2001) are
displayed in Fig.\ 17 which shows O/Fe, Mg/Fe, Si/Fe and Ca/Fe ratios. 
Three
types of symbols are given, circles, squares and dots. The squares show
observations of low metallicity stars with a scatter at very low 
metallicities, as expected if mixing of ejecta with the ISM is not
instantaneous and contributions from individual supernovae with different
progenitor masses give their different signatures. This scatter apparently
becomes as small as the observational uncertainties at [Fe/H]=-2, where
everywhere in the ISM the average signature of SNe II yields
emerges due to an integration over the progenitor mass distribution
(initial mass function, IMF). The open circles indicate the x/Fe and Fe/H 
ratios in a volume of
$5\times 10^4$M$_\odot$ of an initially pristine (big bang) composition,
polluted by a single supernova with our ejecta compositions 
(Thielemann et al.\ 1996, Nomoto et al.\ 1997). 
The numbers inside the
circles refer to the progenitor mass of a supernova whose ejecta composition
was mixed with $5\times 10^4$M$_\odot$ of prestine ISM.
Such a remnant would therefore show directly the [Fe/H] metallicity caused by 
a single SN event. However, we compare with observations from low metallicity 
stars rather than 
remnants. The model stars are indicated by dots (Argast et al.\ 2000, 2001). 
They 
assume a stochastic star formation
at random positions in the ISM and with random progenitor masses (however,
with
a statistical distribution according to a Salpeter IMF). One sees on the one
hand that much smaller contributions to a stellar progenitor can be
possible than expected from a single remnant (leading to Fe/H ratios much
smaller than given in the circles). On the other hand, successive enrichment
of many (and finally overlapping) remnants leads to a metallicity evolution
which approaches at [Fe/H]=-2 the IMF-averaged SNe II yields. These are
apparently in agreement with the observations.

The features (at [Fe/H]=-2) agree nicely with previous ga\-lactic evolution 
calculations which applied our yields 
(Tsujimoto et al.\ 1995, Thomas et al.\ 1998, Matteucci et al.\ 1999, Chiappini et 
al. 1999),  
assuming instantaneous mixing of yields with the ISM. However, the observed
and predicted scatter at very low metallicities bears information which
was previously unavailable. (This approach relies on the assumption that
these difficult low metallicity observations are correct.) The scatter for Si 
and Ca/Fe seems
essentially correct, possibly only slightly too large, which could indicate
that the lower mass SNe II (11 and 13 M$_\odot$) produce slightly too
small ratios and the high mass end (50 M$_\odot$) slightly too large ones.
This can be related to the uncertainty of the mass cut and the assumptions
made on the Fe-yields as a function of progenitor
mass. 

Thus, while the average x/Fe is correct, the progenitor mass dependence
of the Fe-yields could be slightly too strong. This is, however, a small
effect for the elements Si and Ca, produced in explosive burning.
Much larger deviations can be seen for O and Mg, which are dominated by 
uncertainties in stellar evolution (not the explosion). Here the lower
mass stars (11, 13, 15 M$_\odot$) predict clearly much too small ratios,
far beyond the observational scatter, while the 50 M$_\odot$ model predicts
slightly too large ones. This indicates a problem in the stellar models
which might improve with more recent calculations  
(Chieffi et al.\ 1998, Umeda et al.\ 2000, Heger et al.\ 2000ab,
Limongi et al.\ 2000) and has to be tested. 
The main conclusion we can draw from these results
is that such investigations can also test individual stellar yields rather
than only IMF integrated samples. This is a large advantage over the very
few data points we have from individual supernova observations.

Such tests seem also very useful for other applications, where one is (i) not
certain about the stellar site of a nucleosynthesis product or (ii) about
contributions from objects with different evolution timescales.
The possible influence of hypernova contributions, objects beyond the
SN II mass scale which lead to a central black hole but still cause an
explosion  (Nakamura et al.\ 2001), and very massive stars with several 100
M$_\odot$, which undergo a complete disruption by nuclear burning
(VMOs, Heger \& Woosley 2002), should be considered in a 
similar way. In the latter case a full understanding of the initial mass
function, describing the formation frequency as  a function of progenitor
mass, is, however, still marginal.

The r-process is an example where the alternative site to supernovae,
neutron star mergers, occur with a much smaller frequency.
In that case, the mixed phase (orrurring for SNe II at about
[Fe/H]=-2)  should be delayed to larger metallicities. In addition, one expects
with the large amounts of r-process ejecta from each occasionally occurring
neutron star merger (Freiburghaus et al.\ 1999b) a much larger scatter than for
the smoothly changing supernova yields as a function of stellar mass. Both effects
are seen in the r/Fe observations (a scatter of almost a factor of
1000 (Sneden et al.\ 2000, Cayrel et al.\ 2001) at low metallicities,
which still amounts to about a factor of 10 at
[Fe/H]=-1.  Combined with the
suppression of abundances below A=130 in low-metallicity stars, this could 
be taken
as supportive features for a fission cycling r-process. This would also apply,
even if some specific type of supernova is responsible.
Similar galactic evolution calculations as presented here
are needed in order to test the expected amount of scatter as a function
of metallicity to give clues on the r-process site.
The implications for possible r-process site(s)
have been discussed in (Freiburghaus et al.\ 1999b, Qian 2000, 
Cameron 2001ab, Rosswog 2001, Thielemann et al.\ 2002, Qian \& Wasserburg 2002).

\end{document}